\documentclass[%
aps,
prl,
reprint,
superscriptaddress,
showpacs,preprintnumbers,
amsmath,amssymb
]{revtex4-2}

\usepackage{graphicx}   
\usepackage{dcolumn}
\usepackage{bm}
\usepackage{hyperref}
\usepackage{braket}
\usepackage{siunitx}
\usepackage[utf8]{inputenc}

\begin{document}

\title{Observation of giant two-level systems in a granular superconductor}

\author{M. Kristen}	
\affiliation{Institute for Quantum Materials and Technology, Karlsruher Institute of Technology, 76131 Karlsruhe, Germany}	
\affiliation{Physikalisches Institut, Karlsruhe Institute of Technology, 76131 Karlsruhe, Germany}
\author{J. N. Voss}	
\author{M. Wildermuth}
\author{A. Bilmes}	
\author{J. Lisenfeld}	
\affiliation{Physikalisches Institut, Karlsruhe Institute of Technology, 76131 Karlsruhe, Germany}
\author{H. Rotzinger}
\email{rotzinger@kit.edu}	
\author{A. V. Ustinov}	
\affiliation{Institute for Quantum Materials and Technology, Karlsruher Institute of Technology, 76131 Karlsruhe, Germany}	
\affiliation{Physikalisches Institut, Karlsruhe Institute of Technology, 76131 Karlsruhe, Germany}

\date{\today}

\begin{abstract}
Disordered thin films are a common choice of material for superconducting, high impedance circuits used in quantum information or particle detector physics. A wide selection of materials with different levels of granularity are available, but, despite low microwave losses being reported for some, the high degree of disorder always implies the presence of intrinsic defects. Prominently, quantum circuits are prone to interact with two-level systems (TLS), typically originating from solid state defects in the dielectric parts of the circuit, like surface oxides or tunneling barriers.
We present an experimental investigation of TLS in granular aluminum thin films under applied mechanical strain and electric fields. The analysis reveals a class of strongly coupled TLS having electric dipole moments up to $30\, \rm e\AA$, an order of magnitude larger than dipole moments commonly reported for solid state defects. Notably, these large dipole moments appear more often in films with a higher resistivity. Our observations shed new light on granular superconductors and may have implications for their usage as a quantum circuit material.  
\end{abstract}

\maketitle

Originally studied with regard to their enhanced critical temperature \cite{Abeles1966}, thin films of disordered superconductors have regained popularity in recent years due to their outstanding high frequency properties, which stem mainly from their low intrinsic charge carrier density \cite{Beloborodov2007}. In these materials, the large kinetic inductance can lead to circuit impedances in the $\mathrm{k}\Omega$ range. This property has proven valuable for many applications, like single photon detectors \cite{Day2003,Monfardini2011,Valenti2019}, superconducting qubits \cite{Grunhaupt2019,Schon2020,Rieger2022}, nanowire devices \cite{Hongisto2012,Voss2021} and high impedance resonators \cite{Samkharadze2016, Rotzinger2017, Grunhaupt2018, Borisov2020, He2021}. 

The disordered structure of the material also favors the presence of defects throughout the entire film. In particular, defects forming coherent two-level systems (TLS) are known to exist in the amorphous surface oxides of superconducting films. Via their electric dipole moment, such TLS can couple to the AC fields of quantum circuits, causing energy relaxation and dephasing \cite{Muller2019}. As a consequence, increasing experimental effort has been undertaken to investigate TLS origin, nature and location. Individual TLS can be observed in superconducting qubits \cite{Simmonds2004, Grabovskij2012, Lisenfeld2015, Lisenfeld2016, Matityahu2017, Bilmes2021} and resonators \cite{Sarabi2015, Brehm2017}, the latter being also used to study TLS ensembles in different materials \cite{Gao2007, Pappas2011, Burnett2014, Goetz2016, Niepce2020}. 

In this letter, we report on experiments with compact microwave resonators fabricated from granular aluminum thin films. The material consists of pure aluminum grains with a diameter of about $4\, \rm{nm}$ embedded in a matrix of amorphous aluminum oxide, which self-assembles during the deposition of aluminum in an oxygen atmosphere \cite{Deutscher1973}. In the growth process, the oxygen partial pressure influences the thickness of the oxide barrier separating the grains, typically $ 1-2 \, \rm{nm}$, and thus the normal state sheet resistance $R_\mathrm{n}$ \cite{Wildermuth2022}. If $R_\mathrm{n}$ is well below the resistance quantum $R_q = 6.45\,\rm{k}\Omega$, granular aluminum is superconducting ($T_\mathrm{c} \sim 1.8 \, K$) and shows, compared to pure aluminum, a reduced charge carrier density as well as a sizeable kinetic sheet inductance $L_\mathrm{kin}=0.18 \hbar R_\mathrm{n} / \rm{(k_B T_c)}$ in the $\rm nH/\square$ range. 

For this study, we have patterned five different films (chips A-E, see Tab. \ref{tab1} for details) into a total of thirteen $\lambda/2$ microstrip resonators. Depending on the sheet inductance, the resonator length was varied up to $500 \, \si{\micro\metre}$ at a constant width of $2\,\si{\micro\metre}$. On chip D and E, the resonators were covered by an additional, insulating layer of granular aluminum ($R_\mathrm{n} \gg R_\mathrm{q}$), in order to compare the contributions of different film surfaces to the dielectric loss.

\begin{figure*}
	\includegraphics[width=\textwidth]{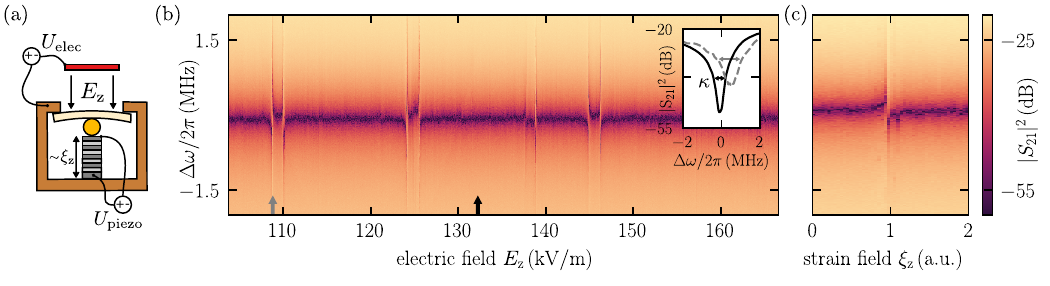}
	\caption{\label{Fig1} Resonator measurements with applied mechanical strain and electric fields. (a) Sketch of the sample holder. A piezo actuator and a DC electrode are used to manipulate the strain $\vec{\xi}$ and the electric field $\vec{E}$. (b) Sample transmission amplitude $|S_{\rm 21}|^2$ in a spectral window $\Delta \omega = \omega - \omega_{\rm r}$ around the resonance frequency at $\omega_{\rm r} = 7.84 \, \mathrm{GHz}$, as a function of the maximum electric field at the film surface $E_\mathrm{z}$. The inset illustrates how the linewidth $\kappa$ of the unperturbed resonance (black) increases near one of the anti-crossings (gray). Inset data is smoothed for visibility. (c) Resonator transmission as a function of mechanical strain applied to the film.}    
\end{figure*}

Figure~\ref{Fig1}(a) provides a sketch of the experimental setup. We apply mechanical strain $\xi_\mathrm{z}$ using a voltage controlled piezo-actuator up to $U_\mathrm{piezo} = 40 \, \mathrm{V}$ (\cite{Grabovskij2012}) and electric fields up to $E_\mathrm{z} = 629 \, \mathrm{kV/m}$ through parallel capacitor plates \cite{Lisenfeld2019}. Here, $E_\mathrm{z}(U_\mathrm{elec})$ is the maximum field value along the resonator film edge, see Supplemental Material (SM) IV for details. To monitor the resonance behavior, the microwave transmission $S_{21}$ of the samples is recorded with a vector network analyzer in a millikelvin temperature setup.

\begin{table}[b]
	\caption{\label{tab1}Overview of the characteristic sample parameters. Each chip hosts several resonators with similar geometry and sheet resistance $R_\mathrm{n}$.}
	\begin{ruledtabular}
		\begin{tabular}{cccccc}
			Chip & A & B & C & D\footnotemark[1] & E\footnotemark[1] \\ 
			Thickness$ \, ({\rm nm})$ & 25 & 22 & 30 & 24+17 & 23+17 \\
			$R_\mathrm{n} \, ({\rm k\Omega/\square})$ & 0.6 & 1.5 & 4.1 & 0.7+12 & 0.7+190 \\
			No. Resonators & 2 & 3 & 3 & 3 &2 \\	
		\end{tabular}
	\end{ruledtabular}
	\footnotetext[1]{Sample is fabricated from a bilayer, which consists of a superconducting (bottom) and insulating (top) granular aluminum film.}
\end{table}

\begin{figure}
	\includegraphics{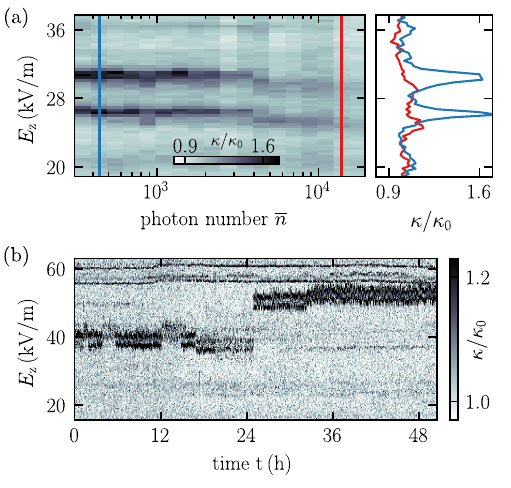}
	\caption{\label{Fig2} Resonator loss traces. (a) Relative resonator linewidth $\kappa/\kappa_{0}$ as a function of the electric field and the average photon number $\overline{n}$. Near an anti-crossing, $\kappa$ is noticeably increased (darker color). The anti-crossings get washed out as the photon number increases $\overline{n} \rightarrow 10^4$. (b) Tracing several anti-crossings over a two-day period. The data reveal various types of temporal fluctuations.}
\end{figure}

For low applied microwave power (few photon regime), the transmission spectrum reveals numerous anti-crossings over a wide range of electric fields, see Fig.~\ref{Fig1}(b) for a subset of the data (strain: Fig.~\ref{Fig1}(c)). In the vicinity of an anti-crossing, the resonance frequency $\omega_\mathrm{r}$ shifts and the linewidth $\kappa$ increases noticeably, as illustrated by the inset of Fig.~\ref{Fig1}(b). A sweep of the power $P_{\rm MW}$, see Fig.~\ref{Fig2}(a), suggests a dependence of this resonance broadening on the average resonator photon number $\overline{n} \propto P_{\rm MW}/(\hbar \omega_\mathrm{r}^2)$ \cite{Schneider2020}. At low photon numbers (blue line), two pronounced peaks in the relative linewidth $\kappa/\kappa_{0}$ clearly indicate an anti-crossing. This feature gradually gets washed out as $\overline{n}$ increases and disappears around $\overline{n} = 10^4$ (red line). Subsequently, we monitored the increased linewidth over a two-day period in a wider range of electric fields (Fig.~\ref{Fig2}(b)). There, the electric field values $E_\mathrm{z}^{i}$, at which the anti-crossings occur, fluctuate noticeably. We identify a telegraphic-like switching pattern between two $E_\mathrm{z}^{i}$ values as well as sudden jumps to a different $E_\mathrm{z}^{i}$.

Overall, the strain and electric field experiments show a similar qualitative picture (see Fig.~\ref{Fig1}(b)+(c)). However, strain measurements are inconsistent due to the hysteretic nature of the piezo actuator, which makes a direct comparison of the data difficult. In the following, we therefore focus on experiments with electric fields, which also show a higher anti-crossing count. 

Under the assumption that the measured anti-crossings stem from the coherent interaction of the resonator with TLS \cite{Muller2019}, the experimental data can be separated into two classes: (1) strongly coupled TLS characterized by a coupling strength $g \approx \kappa$ on the order of the resonator linewidth and (2) moderately coupled TLS with $g \ll \kappa$. While the TLS type (1) generates more pronounced anti-crossings, the TLS of type (2) actually make up most of the total anti-crossings count (SM II+III). 

Due to their typically low lifetimes \cite{Shalibo2010, Lisenfeld2016, Matityahu2017, Lisenfeld2019}, near resonant TLS present a very effective loss channel for superconducting resonators. Since a single TLS can only absorb one photon at a time, the loss is expected to saturate when the number of available photons $\overline{n}$ is large enough such that the TLS are, on average, already excited. This behavior can be seen in Fig.~\ref{Fig2}(a). The telegraphic switching as well as the sudden jumps of the resonator frequency (Fig.~\ref{Fig2}(b)) are also known signatures of TLS dynamics \cite{Burnett2014, Faoro2015, Muller2015, Schlor2019}. Such fluctuations are generally attributed to the coupling of a resonant TLS to secondary TLS at much lower frequencies ($\hbar \omega_\mathrm{TLS} \leq k_\mathrm{B} T$), undergoing incoherent tunneling or random thermal transitions \cite{Faoro2012}. The collective behaviour of the anticrossing-pair around $E_\mathrm{z}^{i}=40\,\mathrm{kV/m}$ indicates that it belongs to a single TLS.

An evaluation of data from all resonators yields an average density of $1/30\,\rm GHz$ for all observable TLS. An independent analysis of the resonator loss tangent $\tan(\delta_0) \propto \kappa(\overline{n} \rightarrow 0)$, however, suggests that these TLS are not solely responsible for the entire dielectric loss, which is more likely to be dominated by a bath of weakly coupled TLS which are not directly observable in our measurements (SM VI).

The quantum mechanical model of a TLS (a tunneling particle in an asymmetric double well potential \cite{Anderson1972, Phillips1972}), yields a transition frequency
\begin{align}
	\omega_{\rm TLS} = \frac{1}{\hbar}\sqrt{\Delta^2+\left(\epsilon + 2 d |E_\mathrm{z}| + 2 \gamma^*\xi_\mathrm{z} \right)^2}.
	\label{Eq0}
\end{align} 
Here, $\Delta$ and $\epsilon$ are the tunnelling and asymmetry energies of the unperturbed TLS, $\gamma^*$ is the effective coupling strength to the strain field $\xi_\mathrm{z} = |\vec{\xi}|$ and $d$ is the component of the TLS' electric dipole moment that is parallel to the maximum expected field strength $E_{\rm z}$. Therefore, $d$ is a lower bound for the TLS dipole moment $|\vec{d}|$. Because of the quadratic terms in Eq.~(\ref{Eq0}) there are always two $E_\mathrm{z}^{\rm ac}$ values where the frequencies of TLS and resonator cross ($\omega_{\rm TLS}=\omega_{\rm r}$), provided that $\Delta < \hbar\omega_{\rm r}$. For moderately coupled TLS with $\Delta \approx \hbar\omega_{\rm r}$, a hyperbolic trace in accordance with Eq.~(\ref{Eq0}) is, on rare occasions, visible in the transmission spectrum over the whole frequency range $\Delta \omega$ (SM II).

\begin{figure*}
	\includegraphics[width=\textwidth]{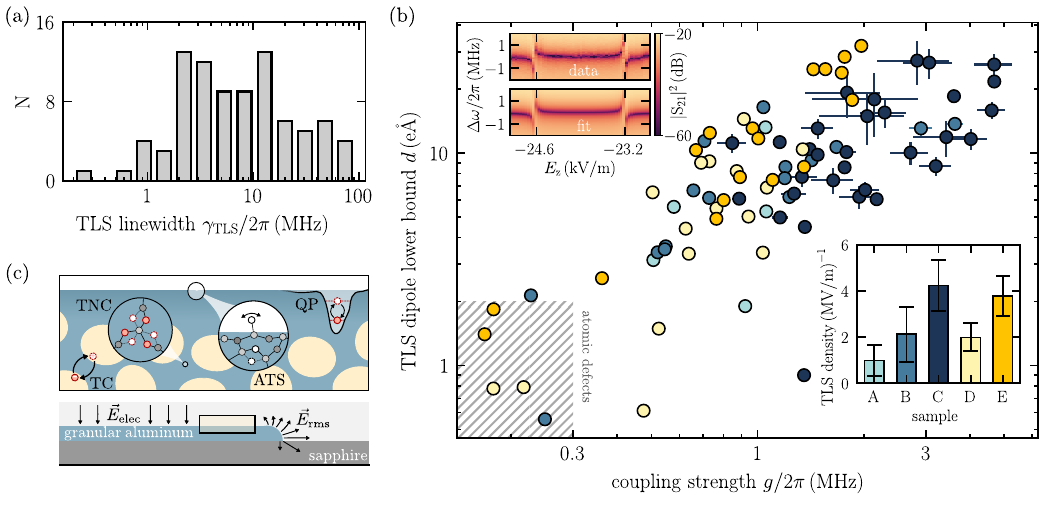}
	\caption{\label{Fig3} Analysis of strongly coupled TLS. (a) Distribution of TLS linewidth $\gamma_{TLS}$ obtained from 86 fits on data from all samples. (b) Resonator-TLS coupling strength $g$ and TLS dipole $d$ extracted from the same fits. Single layer samples are colored in shades of blue, two-layer samples in shades of yellow. Error bars smaller than the marker size are not shown. Hatched area indicates the parameter space expected for atomic defects. Top inset: Transmission spectrum of the resonator in the vicinity of a strongly coupled TLS. Bottom inset: Bottom inset: Average number of strongly coupled TLS in one megavolt per meter electric field range, as found on each sample. Error bars indicate variations between different resonators and cooldowns. (c) Top: Sketch of the granular aluminum film surface, illustrating atomic defects (AD, blue) as well as potential candidates with larger dipole moments (red). The latter includes charges trapped on or between grains (TC), tunneling nanoclusters of atoms (TNC) or quasiparticles trapped due to spatial variations of the superconducting gap (QP). Bottom: Side view of the sample showing the relevant electric fields (not to scale). }
\end{figure*}

We now take a closer look at the pronounced anti-crossings in our spectra. Following Sarabi et al. \cite{Sarabi2015}, the Jaynes-Cummings model for a resonator and a single, strongly coupled TLS yields an analytical expression for the microwave transmission in the vicinity of an anti-crossing
\begin{multline}
S_{21} \propto 1- \\ \frac{\overline{\kappa}_{\rm c}/2}{\kappa/2+i(\omega-\omega_{\rm r})+g^2 (\gamma_{TLS}/2+i(\omega-\omega_{TLS}))^{-1}}.
	\label{Eq2}
\end{multline}

Here, $\kappa$ and $\gamma_{TLS}$ are the linewidths of resonator and TLS. The external loss rate of the resonator $\overline{\kappa}_{\rm c}$ has a complex component due to the non-ideal transmission line \cite{Khalil2012}. We use Eq.~(\ref{Eq2}) together with Eq.~(\ref{Eq0}) to fit the model to the measured transmission data in the vicinity of anti-crossings with strongly coupled TLS. In particular, only symmetric anti-crossing pairs were selected (top inset Fig.~\ref{Fig3}(b)), ensuring unambiguous values for the extracted parameters. On the basis of 86 analyzed anti-crossings obtained in two independent measurement runs, we find a wide distribution of TLS linewidths ranging from 0.2 to $80\, \rm MHz$ (Fig.~\ref{Fig3}(a)). The average linewidth value $\overline{\gamma}_\mathrm{TLS}/2\pi \approx \SI{14}{\mega\hertz}$, corresponding to a coherence time of a few hundreds of nanoseconds, is comparable to values found in similar studies on superconducting qubits \cite{Shalibo2010, Lisenfeld2019} and distributed microwave resonators \cite{Sarabi2014, Brehm2017}. 

The extracted values for the coupling strength $g$ and dipole moment $d$ are shown in Fig.~\ref{Fig3}(b). For reference, the hatched area indicates the estimated parameter space for conventional TLS originating, e.g, from atomic defects (AD), in agreement with earlier experiments that have reported dipoles $|d_\mathrm{AD}|$ of up to $2\, \rm e\AA$\cite{VonSchickfus1977, Golding1979, Martinis2005, Sarabi2016, Lisenfeld2016, Bilmes2020}. While the TLS of type (2) fall into this region, the analysis of the more pronounced anti-crossings shows a different picture. Specifically, order of magnitude larger dipole moments, up to $30 \, \rm e\AA$, and coupling strength $g/2\pi > 1 \, \rm MHz$ are extracted from the fits. Both quantities are statistically correlated (Pearson correlation coefficient $r=0.61$), as expected for $g \propto E_{\rm rms} |\vec{d}|$. Here, $E_{\rm rms} \propto \sqrt{Z_\mathrm{r}}$ is the maximum single-mode resonator field at the film surface (SM IV), which is enhanced by the high characteristic impedance of the resonator $Z_\mathrm{r}$. 

Interestingly, TLS with a coupling strength of $g/2\pi> 2\, \rm MHz$ are only observed on samples B and C. This supports the assumption that for the two-layer samples (D+E), most strongly coupled TLS reside in the insulating top layer, while the resonator currents flow only in the lower, superconducting layer. Then, screening effects partially reduce $E_{\rm rms}$ for these TLS, while $E_\mathrm{z}$ remains comparable for all samples. 

The bottom inset in Fig.~\ref{Fig3}(b) shows the density of strongly coupled TLS found in each sample, normalized to an electric field tuning range of one megavolt per meter. The values increase with $R_\mathrm{n}$, also for the films with the insulating capping layer. In particular, the TLS density in sample E is noticeable higher than in sample A, despite an almost identical sheet resistance of the resonating layer.

Using atomic force microscopy, we measure a rather low surface roughness of $S_\mathrm{q}\leq1$\,nm for the granular aluminum film, which is comparable with the surface roughness of pure aluminum $S_\mathrm{q}\geq1$\,nm. This supports the assumption that the electric fields at the resonator film edge do not experience a substantial enhancement due to the granularity of the film (see also SM IV). Consequently, the observation of dipole moments far exceeding $2\, \rm e\AA$ points towards the existence of TLS having a previously unknown microscopic origin. Possible candidates are illustrated in Fig.~\ref{Fig3}(c). 

In a recent publication, dipole moments of similar size have been observed by studying the dielectric loss of amorphous silicon under a swept electric field \cite{Yu2022}. The findings were interpreted in the scope of a two-TLS model \cite{Schechter2013}. There, TLS are classified, similar to this work, into a weakly and a strongly interacting variety, where the latter can have very large electric dipole moments but is generally only observed under non-equilibrium conditions. In the equilibrium case, phonon mediated TLS-TLS interactions create a gap in the energy spectrum at the relevant energies \cite{Efros1975, Churkin2014}. If, however, the sample is disordered, it has been shown that the granularity modifies the phonon spectrum \cite{Sidorova2022}, which might obstruct the TLS-TLS interaction and lift the gap. 

For the above mechanism to be effective, such TLS have to sit at the grain boundaries rather than the surface oxide of the film. They could, for instance, be formed by tunneling nanoclusters (TNC) with up to hundreds of participating atoms \cite{Lubchenko2001}. For our study, the electrical field has to reach the TLS in order to modify its frequency. As a lower bound, we can roughly estimate the static field penetration depth $\lambda > \lambda_{\rm TF} \simeq 2 \, \rm nm$ to be on the order of the Thomas-Fermi length (SM II). Thus, since $\lambda$ is comparable to the thickness of the films, we can assume that the global field $E_\mathrm{z}$ does not only penetrate the surface oxide but also into the inner film parts, e.g., the oxide between the grains.

The prevalence of strongly coupled TLS in films with higher sheet resistance (inset Fig.~\ref{Fig3}(b)) suggests that the underlying physics might be related to the suppression of the global phase coherent state in these films \cite{Bertrand2019}, which is much more sensitive to changes in $R_\textrm{n}$ than the film morphology. This assumption is supported by noise measurements performed earlier on these samples, where a dependence on $R_\mathrm{n}$ was also observed \cite{Kristen2023}.

A potential TLS candidate that becomes more likely as the film resistance increases are quasiparticles (QP) localized due to spatial variations of the superconducting gap $\delta \Delta_\mathrm{sc}$. This hypothesis has recently been reported as a novel type of TLS in disordered NbN films \cite{deGraaf2020}. Such traps can arise, e.g., due to weak magnetic impurities or film inhomogeneities \cite{Bachar2015, Yang2020}. Compared to NbN, granular aluminum is characterized by an increased coherence length $\xi_\mathrm{0,grAl} \approx 2 \xi_\mathrm{0,NbN}$. The consequence are wider ($\sim \xi_0^2$) traps and thus potentially larger dipoles, which are also shielded less efficiently due to the reduced charge carrier density. 
In granular aluminum, traps can also be expected to be shallower ($\delta \Delta_\mathrm{sc} \approx 0.15\Delta_\mathrm{sc}= 500 \, \rm mK$), resulting in a comparable distribution of TLS frequencies $\propto \Delta_\mathrm{sc}\xi_0^2$ in both materials. We note that the highly-restive layers covering the resonators on D and E might contain superconducting puddles where this effect can also occur \cite{Efetov1980, Pracht2016, Humbert2021}.

Historically, a popular concept to explain strongly coupled TLS are tunneling charges (TC), for which evidence has been found in Josephson junctions \cite{Koch2007a, Lutchyn2008} and Nb+Pt resonators \cite{Burnett2016}. In granular aluminum films, localized charges naturally occur when the sample approaches the superconductor to insulator transition (SIT) \cite{Efetov1980}. While the typical charging energy of fairly isolated grains on the order of several Kelvin does not match the presented findings, it has been shown that hopping charges can be dressed by phononic states of neighboring grains \cite{Agarwal2013} or virtual tunneling processes \cite{Chakravarty1987}, renormalizing the TLS energy.

The localization of charges near the SIT is accompanied by local fluctuations of the superconducting phase \cite{Raychaudhuri2021}. As a consequence, collective excitations of the condensate are expected to acquire dipole moments and appear as a low-energy subgap features \cite{Khvalyuk2023a, Pracht2017}, which would be experimentally observable.

In conclusion, we have characterized two-level systems in oxidized granular aluminum films using electric field and strain tuning experiments. In addition to TLS with conventional properties, we observe a strongly coupled variety of TLS with orders of magnitude larger dipole moments. These TLS are found more frequently in samples with higher sheet resistances. Due to their pronounced frequency fluctuations and low lifetimes, they can be a severe source of noise and dissipation. Since the microscopic mechanism forming these large dipoles remains unknown, we are calling for additional experiments to uncover their nature. Possible pathways include lumped-element resonator geometries \cite{Sarabi2015}, which allow for a more precise investigation of insulating granular aluminum films, applying magnetic field, which offers a selective way to manipulate trapped quasiparticles \cite{deGraaf2020}, measurements with a fast electric field as presented in Ref.~\cite{Yu2022}, or cold-grown granular aluminum films, which have a slightly different grain size distribution \cite{Deutscher1973a}.

The authors thank M. Feigel'man, J. Cole and M.Schechter for helpful discussions and L. Radtke for technical support. Samples were fabricated in the KIT Nanostructure Service Laboratory (NSL). This work was supported by the German Federal Ministry of Education and Research (GeQCoS and QSolid). The authors acknowledge partial support from the Landesgraduiertenf\"orderung of the state Baden-W\"urttemberg (M.W.) and the Helmholtz International Research School for Teratronics (J.N.V.).

\bibliographystyle{apsrev4-2}
\bibliography{bib}

\begin{thebibliography}{74}%
\makeatletter
\providecommand \@ifxundefined [1]{%
 \@ifx{#1\undefined}
}%
\providecommand \@ifnum [1]{%
 \ifnum #1\expandafter \@firstoftwo
 \else \expandafter \@secondoftwo
 \fi
}%
\providecommand \@ifx [1]{%
 \ifx #1\expandafter \@firstoftwo
 \else \expandafter \@secondoftwo
 \fi
}%
\providecommand \natexlab [1]{#1}%
\providecommand \enquote  [1]{``#1''}%
\providecommand \bibnamefont  [1]{#1}%
\providecommand \bibfnamefont [1]{#1}%
\providecommand \citenamefont [1]{#1}%
\providecommand \href@noop [0]{\@secondoftwo}%
\providecommand \href [0]{\begingroup \@sanitize@url \@href}%
\providecommand \@href[1]{\@@startlink{#1}\@@href}%
\providecommand \@@href[1]{\endgroup#1\@@endlink}%
\providecommand \@sanitize@url [0]{\catcode `\\12\catcode `\$12\catcode
  `\&12\catcode `\#12\catcode `\^12\catcode `\_12\catcode `\%12\relax}%
\providecommand \@@startlink[1]{}%
\providecommand \@@endlink[0]{}%
\providecommand \url  [0]{\begingroup\@sanitize@url \@url }%
\providecommand \@url [1]{\endgroup\@href {#1}{\urlprefix }}%
\providecommand \urlprefix  [0]{URL }%
\providecommand \Eprint [0]{\href }%
\providecommand \doibase [0]{https://doi.org/}%
\providecommand \selectlanguage [0]{\@gobble}%
\providecommand \bibinfo  [0]{\@secondoftwo}%
\providecommand \bibfield  [0]{\@secondoftwo}%
\providecommand \translation [1]{[#1]}%
\providecommand \BibitemOpen [0]{}%
\providecommand \bibitemStop [0]{}%
\providecommand \bibitemNoStop [0]{.\EOS\space}%
\providecommand \EOS [0]{\spacefactor3000\relax}%
\providecommand \BibitemShut  [1]{\csname bibitem#1\endcsname}%
\let\auto@bib@innerbib\@empty
\bibitem [{\citenamefont {Abeles}\ \emph {et~al.}(1966)\citenamefont {Abeles},
  \citenamefont {Cohen},\ and\ \citenamefont {Cullen}}]{Abeles1966}%
  \BibitemOpen
  \bibfield  {author} {\bibinfo {author} {\bibfnamefont {B.}~\bibnamefont
  {Abeles}}, \bibinfo {author} {\bibfnamefont {R.~W.}\ \bibnamefont {Cohen}},\
  and\ \bibinfo {author} {\bibfnamefont {G.~W.}\ \bibnamefont {Cullen}},\
  }\href {https://doi.org/10.1103/PhysRevLett.17.632} {\bibfield  {journal}
  {\bibinfo  {journal} {Phys. Rev. Lett.}\ }\textbf {\bibinfo {volume} {17}},\
  \bibinfo {pages} {632} (\bibinfo {year} {1966})}\BibitemShut {NoStop}%
\bibitem [{\citenamefont {Beloborodov}\ \emph {et~al.}(2007)\citenamefont
  {Beloborodov}, \citenamefont {Lopatin}, \citenamefont {Vinokur},\ and\
  \citenamefont {Efetov}}]{Beloborodov2007}%
  \BibitemOpen
  \bibfield  {author} {\bibinfo {author} {\bibfnamefont {I.~S.}\ \bibnamefont
  {Beloborodov}}, \bibinfo {author} {\bibfnamefont {A.~V.}\ \bibnamefont
  {Lopatin}}, \bibinfo {author} {\bibfnamefont {V.~M.}\ \bibnamefont
  {Vinokur}},\ and\ \bibinfo {author} {\bibfnamefont {K.~B.}\ \bibnamefont
  {Efetov}},\ }\href {https://doi.org/10.1103/RevModPhys.79.469} {\bibfield
  {journal} {\bibinfo  {journal} {Rev. Mod. Phys.}\ }\textbf {\bibinfo {volume}
  {79}},\ \bibinfo {pages} {469} (\bibinfo {year} {2007})}\BibitemShut
  {NoStop}%
\bibitem [{\citenamefont {Day}\ \emph {et~al.}(2003)\citenamefont {Day},
  \citenamefont {LeDuc}, \citenamefont {Mazin}, \citenamefont {Vayonakis},\
  and\ \citenamefont {Zmuidzinas}}]{Day2003}%
  \BibitemOpen
  \bibfield  {author} {\bibinfo {author} {\bibfnamefont {P.~K.}\ \bibnamefont
  {Day}}, \bibinfo {author} {\bibfnamefont {H.~G.}\ \bibnamefont {LeDuc}},
  \bibinfo {author} {\bibfnamefont {B.~A.}\ \bibnamefont {Mazin}}, \bibinfo
  {author} {\bibfnamefont {A.}~\bibnamefont {Vayonakis}},\ and\ \bibinfo
  {author} {\bibfnamefont {J.}~\bibnamefont {Zmuidzinas}},\ }\href
  {https://doi.org/10.1038/nature02037} {\bibfield  {journal} {\bibinfo
  {journal} {Nature}\ }\textbf {\bibinfo {volume} {425}},\ \bibinfo {pages}
  {817} (\bibinfo {year} {2003})}\BibitemShut {NoStop}%
\bibitem [{\citenamefont {Monfardini}\ \emph {et~al.}(2011)\citenamefont
  {Monfardini}, \citenamefont {Benoit}, \citenamefont {Bideaud}, \citenamefont
  {Swenson}, \citenamefont {Cruciani}, \citenamefont {Camus}, \citenamefont
  {Hoffmann}, \citenamefont {Désert}, \citenamefont {Doyle}, \citenamefont
  {Ade}, \citenamefont {Mauskopf}, \citenamefont {Tucker}, \citenamefont
  {Roesch}, \citenamefont {Leclercq}, \citenamefont {Schuster}, \citenamefont
  {Endo}, \citenamefont {Baryshev}, \citenamefont {Baselmans}, \citenamefont
  {Ferrari}, \citenamefont {Yates}, \citenamefont {Bourrion}, \citenamefont
  {Macias-Perez}, \citenamefont {Vescovi}, \citenamefont {Calvo},\ and\
  \citenamefont {Giordano}}]{Monfardini2011}%
  \BibitemOpen
  \bibfield  {author} {\bibinfo {author} {\bibfnamefont {A.}~\bibnamefont
  {Monfardini}}, \bibinfo {author} {\bibfnamefont {A.}~\bibnamefont {Benoit}},
  \bibinfo {author} {\bibfnamefont {A.}~\bibnamefont {Bideaud}}, \bibinfo
  {author} {\bibfnamefont {L.}~\bibnamefont {Swenson}}, \bibinfo {author}
  {\bibfnamefont {A.}~\bibnamefont {Cruciani}}, \bibinfo {author}
  {\bibfnamefont {P.}~\bibnamefont {Camus}}, \bibinfo {author} {\bibfnamefont
  {C.}~\bibnamefont {Hoffmann}}, \bibinfo {author} {\bibfnamefont {F.~X.}\
  \bibnamefont {Désert}}, \bibinfo {author} {\bibfnamefont {S.}~\bibnamefont
  {Doyle}}, \bibinfo {author} {\bibfnamefont {P.}~\bibnamefont {Ade}}, \bibinfo
  {author} {\bibfnamefont {P.}~\bibnamefont {Mauskopf}}, \bibinfo {author}
  {\bibfnamefont {C.}~\bibnamefont {Tucker}}, \bibinfo {author} {\bibfnamefont
  {M.}~\bibnamefont {Roesch}}, \bibinfo {author} {\bibfnamefont
  {S.}~\bibnamefont {Leclercq}}, \bibinfo {author} {\bibfnamefont {K.~F.}\
  \bibnamefont {Schuster}}, \bibinfo {author} {\bibfnamefont {A.}~\bibnamefont
  {Endo}}, \bibinfo {author} {\bibfnamefont {A.}~\bibnamefont {Baryshev}},
  \bibinfo {author} {\bibfnamefont {J.~J.~A.}\ \bibnamefont {Baselmans}},
  \bibinfo {author} {\bibfnamefont {L.}~\bibnamefont {Ferrari}}, \bibinfo
  {author} {\bibfnamefont {S.~J.~C.}\ \bibnamefont {Yates}}, \bibinfo {author}
  {\bibfnamefont {O.}~\bibnamefont {Bourrion}}, \bibinfo {author}
  {\bibfnamefont {J.}~\bibnamefont {Macias-Perez}}, \bibinfo {author}
  {\bibfnamefont {C.}~\bibnamefont {Vescovi}}, \bibinfo {author} {\bibfnamefont
  {M.}~\bibnamefont {Calvo}},\ and\ \bibinfo {author} {\bibfnamefont
  {C.}~\bibnamefont {Giordano}},\ }\href
  {https://doi.org/10.1088/0067-0049/194/2/24} {\bibfield  {journal} {\bibinfo
  {journal} {ApJS}\ }\textbf {\bibinfo {volume} {194}},\ \bibinfo {pages} {24}
  (\bibinfo {year} {2011})}\BibitemShut {NoStop}%
\bibitem [{\citenamefont {Valenti}\ \emph {et~al.}(2019)\citenamefont
  {Valenti}, \citenamefont {Henriques}, \citenamefont {Catelani}, \citenamefont
  {Maleeva}, \citenamefont {Grünhaupt}, \citenamefont {von Lüpke},
  \citenamefont {Skacel}, \citenamefont {Winkel}, \citenamefont {Bilmes},
  \citenamefont {Ustinov}, \citenamefont {Goupy}, \citenamefont {Calvo},
  \citenamefont {Benoît}, \citenamefont {Levy-Bertrand}, \citenamefont
  {Monfardini},\ and\ \citenamefont {Pop}}]{Valenti2019}%
  \BibitemOpen
  \bibfield  {author} {\bibinfo {author} {\bibfnamefont {F.}~\bibnamefont
  {Valenti}}, \bibinfo {author} {\bibfnamefont {F.}~\bibnamefont {Henriques}},
  \bibinfo {author} {\bibfnamefont {G.}~\bibnamefont {Catelani}}, \bibinfo
  {author} {\bibfnamefont {N.}~\bibnamefont {Maleeva}}, \bibinfo {author}
  {\bibfnamefont {L.}~\bibnamefont {Grünhaupt}}, \bibinfo {author}
  {\bibfnamefont {U.}~\bibnamefont {von Lüpke}}, \bibinfo {author}
  {\bibfnamefont {S.~T.}\ \bibnamefont {Skacel}}, \bibinfo {author}
  {\bibfnamefont {P.}~\bibnamefont {Winkel}}, \bibinfo {author} {\bibfnamefont
  {A.}~\bibnamefont {Bilmes}}, \bibinfo {author} {\bibfnamefont {A.~V.}\
  \bibnamefont {Ustinov}}, \bibinfo {author} {\bibfnamefont {J.}~\bibnamefont
  {Goupy}}, \bibinfo {author} {\bibfnamefont {M.}~\bibnamefont {Calvo}},
  \bibinfo {author} {\bibfnamefont {A.}~\bibnamefont {Benoît}}, \bibinfo
  {author} {\bibfnamefont {F.}~\bibnamefont {Levy-Bertrand}}, \bibinfo {author}
  {\bibfnamefont {A.}~\bibnamefont {Monfardini}},\ and\ \bibinfo {author}
  {\bibfnamefont {I.~M.}\ \bibnamefont {Pop}},\ }\href
  {https://doi.org/10.1103/PhysRevApplied.11.054087} {\bibfield  {journal}
  {\bibinfo  {journal} {Physical Review Applied}\ }\textbf {\bibinfo {volume}
  {11}},\ \bibinfo {pages} {054087} (\bibinfo {year} {2019})}\BibitemShut
  {NoStop}%
\bibitem [{\citenamefont {Grünhaupt}\ \emph {et~al.}(2019)\citenamefont
  {Grünhaupt}, \citenamefont {Spiecker}, \citenamefont {Gusenkova},
  \citenamefont {Maleeva}, \citenamefont {Skacel}, \citenamefont {Takmakov},
  \citenamefont {Valenti}, \citenamefont {Winkel}, \citenamefont {Rotzinger},
  \citenamefont {Wernsdorfer}, \citenamefont {Ustinov},\ and\ \citenamefont
  {Pop}}]{Grunhaupt2019}%
  \BibitemOpen
  \bibfield  {author} {\bibinfo {author} {\bibfnamefont {L.}~\bibnamefont
  {Grünhaupt}}, \bibinfo {author} {\bibfnamefont {M.}~\bibnamefont
  {Spiecker}}, \bibinfo {author} {\bibfnamefont {D.}~\bibnamefont {Gusenkova}},
  \bibinfo {author} {\bibfnamefont {N.}~\bibnamefont {Maleeva}}, \bibinfo
  {author} {\bibfnamefont {S.~T.}\ \bibnamefont {Skacel}}, \bibinfo {author}
  {\bibfnamefont {I.}~\bibnamefont {Takmakov}}, \bibinfo {author}
  {\bibfnamefont {F.}~\bibnamefont {Valenti}}, \bibinfo {author} {\bibfnamefont
  {P.}~\bibnamefont {Winkel}}, \bibinfo {author} {\bibfnamefont
  {H.}~\bibnamefont {Rotzinger}}, \bibinfo {author} {\bibfnamefont
  {W.}~\bibnamefont {Wernsdorfer}}, \bibinfo {author} {\bibfnamefont {A.~V.}\
  \bibnamefont {Ustinov}},\ and\ \bibinfo {author} {\bibfnamefont {I.~M.}\
  \bibnamefont {Pop}},\ }\href {https://doi.org/10.1038/s41563-019-0350-3}
  {\bibfield  {journal} {\bibinfo  {journal} {Nature Materials}\ }\textbf
  {\bibinfo {volume} {18}},\ \bibinfo {pages} {816} (\bibinfo {year}
  {2019})}\BibitemShut {NoStop}%
\bibitem [{\citenamefont {Schön}\ \emph {et~al.}(2020)\citenamefont {Schön},
  \citenamefont {Voss}, \citenamefont {Wildermuth}, \citenamefont {Schneider},
  \citenamefont {Skacel}, \citenamefont {Weides}, \citenamefont {Cole},
  \citenamefont {Rotzinger},\ and\ \citenamefont {Ustinov}}]{Schon2020}%
  \BibitemOpen
  \bibfield  {author} {\bibinfo {author} {\bibfnamefont {Y.}~\bibnamefont
  {Schön}}, \bibinfo {author} {\bibfnamefont {J.~N.}\ \bibnamefont {Voss}},
  \bibinfo {author} {\bibfnamefont {M.}~\bibnamefont {Wildermuth}}, \bibinfo
  {author} {\bibfnamefont {A.}~\bibnamefont {Schneider}}, \bibinfo {author}
  {\bibfnamefont {S.~T.}\ \bibnamefont {Skacel}}, \bibinfo {author}
  {\bibfnamefont {M.~P.}\ \bibnamefont {Weides}}, \bibinfo {author}
  {\bibfnamefont {J.~H.}\ \bibnamefont {Cole}}, \bibinfo {author}
  {\bibfnamefont {H.}~\bibnamefont {Rotzinger}},\ and\ \bibinfo {author}
  {\bibfnamefont {A.~V.}\ \bibnamefont {Ustinov}},\ }\href
  {https://doi.org/10.1038/s41535-020-0220-x} {\bibfield  {journal} {\bibinfo
  {journal} {npj Quantum Mater.}\ }\textbf {\bibinfo {volume} {5}},\ \bibinfo
  {pages} {1} (\bibinfo {year} {2020})}\BibitemShut {NoStop}%
\bibitem [{\citenamefont {Rieger}\ \emph {et~al.}(2022)\citenamefont {Rieger},
  \citenamefont {Günzler}, \citenamefont {Spiecker}, \citenamefont {Paluch},
  \citenamefont {Winkel}, \citenamefont {Hahn}, \citenamefont {Hohmann},
  \citenamefont {Bacher}, \citenamefont {Wernsdorfer},\ and\ \citenamefont
  {Pop}}]{Rieger2022}%
  \BibitemOpen
  \bibfield  {author} {\bibinfo {author} {\bibfnamefont {D.}~\bibnamefont
  {Rieger}}, \bibinfo {author} {\bibfnamefont {S.}~\bibnamefont {Günzler}},
  \bibinfo {author} {\bibfnamefont {M.}~\bibnamefont {Spiecker}}, \bibinfo
  {author} {\bibfnamefont {P.}~\bibnamefont {Paluch}}, \bibinfo {author}
  {\bibfnamefont {P.}~\bibnamefont {Winkel}}, \bibinfo {author} {\bibfnamefont
  {L.}~\bibnamefont {Hahn}}, \bibinfo {author} {\bibfnamefont {J.~K.}\
  \bibnamefont {Hohmann}}, \bibinfo {author} {\bibfnamefont {A.}~\bibnamefont
  {Bacher}}, \bibinfo {author} {\bibfnamefont {W.}~\bibnamefont
  {Wernsdorfer}},\ and\ \bibinfo {author} {\bibfnamefont {I.~M.}\ \bibnamefont
  {Pop}},\ }\href {https://doi.org/10.1038/s41563-022-01417-9} {\bibfield
  {journal} {\bibinfo  {journal} {Nat. Mater.}\ ,\ \bibinfo {pages} {1}}
  (\bibinfo {year} {2022})}\BibitemShut {NoStop}%
\bibitem [{\citenamefont {Hongisto}\ and\ \citenamefont
  {Zorin}(2012)}]{Hongisto2012}%
  \BibitemOpen
  \bibfield  {author} {\bibinfo {author} {\bibfnamefont {T.~T.}\ \bibnamefont
  {Hongisto}}\ and\ \bibinfo {author} {\bibfnamefont {A.~B.}\ \bibnamefont
  {Zorin}},\ }\href {https://doi.org/10.1103/PhysRevLett.108.097001} {\bibfield
   {journal} {\bibinfo  {journal} {Phys. Rev. Lett.}\ }\textbf {\bibinfo
  {volume} {108}},\ \bibinfo {pages} {097001} (\bibinfo {year}
  {2012})}\BibitemShut {NoStop}%
\bibitem [{\citenamefont {Voss}\ \emph {et~al.}(2021)\citenamefont {Voss},
  \citenamefont {Schön}, \citenamefont {Wildermuth}, \citenamefont {Dorer},
  \citenamefont {Cole}, \citenamefont {Rotzinger},\ and\ \citenamefont
  {Ustinov}}]{Voss2021}%
  \BibitemOpen
  \bibfield  {author} {\bibinfo {author} {\bibfnamefont {J.~N.}\ \bibnamefont
  {Voss}}, \bibinfo {author} {\bibfnamefont {Y.}~\bibnamefont {Schön}},
  \bibinfo {author} {\bibfnamefont {M.}~\bibnamefont {Wildermuth}}, \bibinfo
  {author} {\bibfnamefont {D.}~\bibnamefont {Dorer}}, \bibinfo {author}
  {\bibfnamefont {J.~H.}\ \bibnamefont {Cole}}, \bibinfo {author}
  {\bibfnamefont {H.}~\bibnamefont {Rotzinger}},\ and\ \bibinfo {author}
  {\bibfnamefont {A.~V.}\ \bibnamefont {Ustinov}},\ }\href
  {https://doi.org/10.1021/acsnano.0c08721} {\bibfield  {journal} {\bibinfo
  {journal} {ACS Nano}\ }\textbf {\bibinfo {volume} {15}},\ \bibinfo {pages}
  {4108} (\bibinfo {year} {2021})}\BibitemShut {NoStop}%
\bibitem [{\citenamefont {Samkharadze}\ \emph {et~al.}(2016)\citenamefont
  {Samkharadze}, \citenamefont {Bruno}, \citenamefont {Scarlino}, \citenamefont
  {Zheng}, \citenamefont {DiVincenzo}, \citenamefont {DiCarlo},\ and\
  \citenamefont {Vandersypen}}]{Samkharadze2016}%
  \BibitemOpen
  \bibfield  {author} {\bibinfo {author} {\bibfnamefont {N.}~\bibnamefont
  {Samkharadze}}, \bibinfo {author} {\bibfnamefont {A.}~\bibnamefont {Bruno}},
  \bibinfo {author} {\bibfnamefont {P.}~\bibnamefont {Scarlino}}, \bibinfo
  {author} {\bibfnamefont {G.}~\bibnamefont {Zheng}}, \bibinfo {author}
  {\bibfnamefont {D.}~\bibnamefont {DiVincenzo}}, \bibinfo {author}
  {\bibfnamefont {L.}~\bibnamefont {DiCarlo}},\ and\ \bibinfo {author}
  {\bibfnamefont {L.}~\bibnamefont {Vandersypen}},\ }\href
  {https://doi.org/10.1103/PhysRevApplied.5.044004} {\bibfield  {journal}
  {\bibinfo  {journal} {Physical Review Applied}\ }\textbf {\bibinfo {volume}
  {5}},\ \bibinfo {pages} {044004} (\bibinfo {year} {2016})}\BibitemShut
  {NoStop}%
\bibitem [{\citenamefont {Rotzinger}\ \emph {et~al.}(2017)\citenamefont
  {Rotzinger}, \citenamefont {Skacel}, \citenamefont {Pfirrmann}, \citenamefont
  {Voss}, \citenamefont {Münzberg}, \citenamefont {Probst}, \citenamefont
  {Bushev}, \citenamefont {Weides}, \citenamefont {Ustinov},\ and\
  \citenamefont {Mooij}}]{Rotzinger2017}%
  \BibitemOpen
  \bibfield  {author} {\bibinfo {author} {\bibfnamefont {H.}~\bibnamefont
  {Rotzinger}}, \bibinfo {author} {\bibfnamefont {S.~T.}\ \bibnamefont
  {Skacel}}, \bibinfo {author} {\bibfnamefont {M.}~\bibnamefont {Pfirrmann}},
  \bibinfo {author} {\bibfnamefont {J.~N.}\ \bibnamefont {Voss}}, \bibinfo
  {author} {\bibfnamefont {J.}~\bibnamefont {Münzberg}}, \bibinfo {author}
  {\bibfnamefont {S.}~\bibnamefont {Probst}}, \bibinfo {author} {\bibfnamefont
  {P.}~\bibnamefont {Bushev}}, \bibinfo {author} {\bibfnamefont {M.~P.}\
  \bibnamefont {Weides}}, \bibinfo {author} {\bibfnamefont {A.~V.}\
  \bibnamefont {Ustinov}},\ and\ \bibinfo {author} {\bibfnamefont {J.~E.}\
  \bibnamefont {Mooij}},\ }\href
  {https://doi.org/10.1088/0953-2048/30/2/025002} {\bibfield  {journal}
  {\bibinfo  {journal} {Supercond. Sci. Technol.}\ }\textbf {\bibinfo {volume}
  {30}},\ \bibinfo {pages} {025002} (\bibinfo {year} {2017})}\BibitemShut
  {NoStop}%
\bibitem [{\citenamefont {Grünhaupt}\ \emph {et~al.}(2018)\citenamefont
  {Grünhaupt}, \citenamefont {Maleeva}, \citenamefont {Skacel}, \citenamefont
  {Calvo}, \citenamefont {Levy-Bertrand}, \citenamefont {Ustinov},
  \citenamefont {Rotzinger}, \citenamefont {Monfardini}, \citenamefont
  {Catelani},\ and\ \citenamefont {Pop}}]{Grunhaupt2018}%
  \BibitemOpen
  \bibfield  {author} {\bibinfo {author} {\bibfnamefont {L.}~\bibnamefont
  {Grünhaupt}}, \bibinfo {author} {\bibfnamefont {N.}~\bibnamefont {Maleeva}},
  \bibinfo {author} {\bibfnamefont {S.~T.}\ \bibnamefont {Skacel}}, \bibinfo
  {author} {\bibfnamefont {M.}~\bibnamefont {Calvo}}, \bibinfo {author}
  {\bibfnamefont {F.}~\bibnamefont {Levy-Bertrand}}, \bibinfo {author}
  {\bibfnamefont {A.~V.}\ \bibnamefont {Ustinov}}, \bibinfo {author}
  {\bibfnamefont {H.}~\bibnamefont {Rotzinger}}, \bibinfo {author}
  {\bibfnamefont {A.}~\bibnamefont {Monfardini}}, \bibinfo {author}
  {\bibfnamefont {G.}~\bibnamefont {Catelani}},\ and\ \bibinfo {author}
  {\bibfnamefont {I.~M.}\ \bibnamefont {Pop}},\ }\href
  {https://doi.org/10.1103/PhysRevLett.121.117001} {\bibfield  {journal}
  {\bibinfo  {journal} {Phys. Rev. Lett.}\ }\textbf {\bibinfo {volume} {121}},\
  \bibinfo {pages} {117001} (\bibinfo {year} {2018})}\BibitemShut {NoStop}%
\bibitem [{\citenamefont {Borisov}\ \emph {et~al.}(2020)\citenamefont
  {Borisov}, \citenamefont {Rieger}, \citenamefont {Winkel}, \citenamefont
  {Henriques}, \citenamefont {Valenti}, \citenamefont {Ionita}, \citenamefont
  {Wessbecher}, \citenamefont {Spiecker}, \citenamefont {Gusenkova},
  \citenamefont {Pop},\ and\ \citenamefont {Wernsdorfer}}]{Borisov2020}%
  \BibitemOpen
  \bibfield  {author} {\bibinfo {author} {\bibfnamefont {K.}~\bibnamefont
  {Borisov}}, \bibinfo {author} {\bibfnamefont {D.}~\bibnamefont {Rieger}},
  \bibinfo {author} {\bibfnamefont {P.}~\bibnamefont {Winkel}}, \bibinfo
  {author} {\bibfnamefont {F.}~\bibnamefont {Henriques}}, \bibinfo {author}
  {\bibfnamefont {F.}~\bibnamefont {Valenti}}, \bibinfo {author} {\bibfnamefont
  {A.}~\bibnamefont {Ionita}}, \bibinfo {author} {\bibfnamefont
  {M.}~\bibnamefont {Wessbecher}}, \bibinfo {author} {\bibfnamefont
  {M.}~\bibnamefont {Spiecker}}, \bibinfo {author} {\bibfnamefont
  {D.}~\bibnamefont {Gusenkova}}, \bibinfo {author} {\bibfnamefont {I.~M.}\
  \bibnamefont {Pop}},\ and\ \bibinfo {author} {\bibfnamefont {W.}~\bibnamefont
  {Wernsdorfer}},\ }\href {https://doi.org/10.1063/5.0018012} {\bibfield
  {journal} {\bibinfo  {journal} {Appl. Phys. Lett.}\ }\textbf {\bibinfo
  {volume} {117}},\ \bibinfo {pages} {120502} (\bibinfo {year}
  {2020})}\BibitemShut {NoStop}%
\bibitem [{\citenamefont {He}\ \emph {et~al.}(2021)\citenamefont {He},
  \citenamefont {OuYang}, \citenamefont {Dai}, \citenamefont {Guan},
  \citenamefont {Hu}, \citenamefont {He}, \citenamefont {Wang},\ and\
  \citenamefont {Wei}}]{He2021}%
  \BibitemOpen
  \bibfield  {author} {\bibinfo {author} {\bibfnamefont {Q.}~\bibnamefont
  {He}}, \bibinfo {author} {\bibfnamefont {P.}~\bibnamefont {OuYang}}, \bibinfo
  {author} {\bibfnamefont {M.}~\bibnamefont {Dai}}, \bibinfo {author}
  {\bibfnamefont {H.}~\bibnamefont {Guan}}, \bibinfo {author} {\bibfnamefont
  {J.}~\bibnamefont {Hu}}, \bibinfo {author} {\bibfnamefont {S.}~\bibnamefont
  {He}}, \bibinfo {author} {\bibfnamefont {Y.}~\bibnamefont {Wang}},\ and\
  \bibinfo {author} {\bibfnamefont {L.~F.}\ \bibnamefont {Wei}},\ }\href
  {https://doi.org/10.1063/5.0046910} {\bibfield  {journal} {\bibinfo
  {journal} {AIP Adv.}\ }\textbf {\bibinfo {volume} {11}},\ \bibinfo {pages}
  {065204} (\bibinfo {year} {2021})}\BibitemShut {NoStop}%
\bibitem [{\citenamefont {Müller}\ \emph {et~al.}(2019)\citenamefont
  {Müller}, \citenamefont {Cole},\ and\ \citenamefont
  {Lisenfeld}}]{Muller2019}%
  \BibitemOpen
  \bibfield  {author} {\bibinfo {author} {\bibfnamefont {C.}~\bibnamefont
  {Müller}}, \bibinfo {author} {\bibfnamefont {J.~H.}\ \bibnamefont {Cole}},\
  and\ \bibinfo {author} {\bibfnamefont {J.}~\bibnamefont {Lisenfeld}},\ }\href
  {https://doi.org/10.1088/1361-6633/ab3a7e} {\bibfield  {journal} {\bibinfo
  {journal} {Rep. Prog. Phys}\ }\textbf {\bibinfo {volume} {82}},\ \bibinfo
  {pages} {31} (\bibinfo {year} {2019})}\BibitemShut {NoStop}%
\bibitem [{\citenamefont {Simmonds}\ \emph {et~al.}(2004)\citenamefont
  {Simmonds}, \citenamefont {Lang}, \citenamefont {Hite}, \citenamefont {Nam},
  \citenamefont {Pappas},\ and\ \citenamefont {Martinis}}]{Simmonds2004}%
  \BibitemOpen
  \bibfield  {author} {\bibinfo {author} {\bibfnamefont {R.~W.}\ \bibnamefont
  {Simmonds}}, \bibinfo {author} {\bibfnamefont {K.~M.}\ \bibnamefont {Lang}},
  \bibinfo {author} {\bibfnamefont {D.~A.}\ \bibnamefont {Hite}}, \bibinfo
  {author} {\bibfnamefont {S.}~\bibnamefont {Nam}}, \bibinfo {author}
  {\bibfnamefont {D.~P.}\ \bibnamefont {Pappas}},\ and\ \bibinfo {author}
  {\bibfnamefont {J.~M.}\ \bibnamefont {Martinis}},\ }\href
  {https://doi.org/10.1103/PhysRevLett.93.077003} {\bibfield  {journal}
  {\bibinfo  {journal} {Phys. Rev. Lett.}\ }\textbf {\bibinfo {volume} {93}},\
  \bibinfo {pages} {077003} (\bibinfo {year} {2004})}\BibitemShut {NoStop}%
\bibitem [{\citenamefont {Grabovskij}\ \emph {et~al.}(2012)\citenamefont
  {Grabovskij}, \citenamefont {Peichl}, \citenamefont {Lisenfeld},
  \citenamefont {Weiss},\ and\ \citenamefont {Ustinov}}]{Grabovskij2012}%
  \BibitemOpen
  \bibfield  {author} {\bibinfo {author} {\bibfnamefont {G.~J.}\ \bibnamefont
  {Grabovskij}}, \bibinfo {author} {\bibfnamefont {T.}~\bibnamefont {Peichl}},
  \bibinfo {author} {\bibfnamefont {J.}~\bibnamefont {Lisenfeld}}, \bibinfo
  {author} {\bibfnamefont {G.}~\bibnamefont {Weiss}},\ and\ \bibinfo {author}
  {\bibfnamefont {A.~V.}\ \bibnamefont {Ustinov}},\ }\href
  {https://doi.org/10.1126/science.1226487} {\bibfield  {journal} {\bibinfo
  {journal} {Science (New York, N.Y.)}\ }\textbf {\bibinfo {volume} {338}},\
  \bibinfo {pages} {232} (\bibinfo {year} {2012})}\BibitemShut {NoStop}%
\bibitem [{\citenamefont {Lisenfeld}\ \emph {et~al.}(2015)\citenamefont
  {Lisenfeld}, \citenamefont {Grabovskij}, \citenamefont {Müller},
  \citenamefont {Cole}, \citenamefont {Weiss},\ and\ \citenamefont
  {Ustinov}}]{Lisenfeld2015}%
  \BibitemOpen
  \bibfield  {author} {\bibinfo {author} {\bibfnamefont {J.}~\bibnamefont
  {Lisenfeld}}, \bibinfo {author} {\bibfnamefont {G.~J.}\ \bibnamefont
  {Grabovskij}}, \bibinfo {author} {\bibfnamefont {C.}~\bibnamefont {Müller}},
  \bibinfo {author} {\bibfnamefont {J.~H.}\ \bibnamefont {Cole}}, \bibinfo
  {author} {\bibfnamefont {G.}~\bibnamefont {Weiss}},\ and\ \bibinfo {author}
  {\bibfnamefont {A.~V.}\ \bibnamefont {Ustinov}},\ }\href
  {https://doi.org/10.1038/ncomms7182} {\bibfield  {journal} {\bibinfo
  {journal} {Nat. Commun.}\ }\textbf {\bibinfo {volume} {6}},\ \bibinfo {pages}
  {6182} (\bibinfo {year} {2015})}\BibitemShut {NoStop}%
\bibitem [{\citenamefont {Lisenfeld}\ \emph {et~al.}(2016)\citenamefont
  {Lisenfeld}, \citenamefont {Bilmes}, \citenamefont {Matityahu}, \citenamefont
  {Zanker}, \citenamefont {Marthaler}, \citenamefont {Schechter}, \citenamefont
  {Schön}, \citenamefont {Shnirman}, \citenamefont {Weiss},\ and\
  \citenamefont {Ustinov}}]{Lisenfeld2016}%
  \BibitemOpen
  \bibfield  {author} {\bibinfo {author} {\bibfnamefont {J.}~\bibnamefont
  {Lisenfeld}}, \bibinfo {author} {\bibfnamefont {A.}~\bibnamefont {Bilmes}},
  \bibinfo {author} {\bibfnamefont {S.}~\bibnamefont {Matityahu}}, \bibinfo
  {author} {\bibfnamefont {S.}~\bibnamefont {Zanker}}, \bibinfo {author}
  {\bibfnamefont {M.}~\bibnamefont {Marthaler}}, \bibinfo {author}
  {\bibfnamefont {M.}~\bibnamefont {Schechter}}, \bibinfo {author}
  {\bibfnamefont {G.}~\bibnamefont {Schön}}, \bibinfo {author} {\bibfnamefont
  {A.}~\bibnamefont {Shnirman}}, \bibinfo {author} {\bibfnamefont
  {G.}~\bibnamefont {Weiss}},\ and\ \bibinfo {author} {\bibfnamefont {A.~V.}\
  \bibnamefont {Ustinov}},\ }\href {https://doi.org/10.1038/srep23786}
  {\bibfield  {journal} {\bibinfo  {journal} {Sci. Rep.}\ }\textbf {\bibinfo
  {volume} {6}},\ \bibinfo {pages} {23786} (\bibinfo {year}
  {2016})}\BibitemShut {NoStop}%
\bibitem [{\citenamefont {Matityahu}\ \emph {et~al.}(2017)\citenamefont
  {Matityahu}, \citenamefont {Lisenfeld}, \citenamefont {Bilmes}, \citenamefont
  {Shnirman}, \citenamefont {Weiss}, \citenamefont {Ustinov},\ and\
  \citenamefont {Schechter}}]{Matityahu2017}%
  \BibitemOpen
  \bibfield  {author} {\bibinfo {author} {\bibfnamefont {S.}~\bibnamefont
  {Matityahu}}, \bibinfo {author} {\bibfnamefont {J.}~\bibnamefont
  {Lisenfeld}}, \bibinfo {author} {\bibfnamefont {A.}~\bibnamefont {Bilmes}},
  \bibinfo {author} {\bibfnamefont {A.}~\bibnamefont {Shnirman}}, \bibinfo
  {author} {\bibfnamefont {G.}~\bibnamefont {Weiss}}, \bibinfo {author}
  {\bibfnamefont {A.~V.}\ \bibnamefont {Ustinov}},\ and\ \bibinfo {author}
  {\bibfnamefont {M.}~\bibnamefont {Schechter}},\ }\href
  {https://doi.org/10.1103/PhysRevB.95.241409} {\bibfield  {journal} {\bibinfo
  {journal} {Phys. Rev. B}\ }\textbf {\bibinfo {volume} {95}},\ \bibinfo
  {pages} {241409} (\bibinfo {year} {2017})}\BibitemShut {NoStop}%
\bibitem [{\citenamefont {Bilmes}\ \emph {et~al.}(2021)\citenamefont {Bilmes},
  \citenamefont {Volosheniuk}, \citenamefont {Brehm}, \citenamefont {Ustinov},\
  and\ \citenamefont {Lisenfeld}}]{Bilmes2021}%
  \BibitemOpen
  \bibfield  {author} {\bibinfo {author} {\bibfnamefont {A.}~\bibnamefont
  {Bilmes}}, \bibinfo {author} {\bibfnamefont {S.}~\bibnamefont {Volosheniuk}},
  \bibinfo {author} {\bibfnamefont {J.~D.}\ \bibnamefont {Brehm}}, \bibinfo
  {author} {\bibfnamefont {A.~V.}\ \bibnamefont {Ustinov}},\ and\ \bibinfo
  {author} {\bibfnamefont {J.}~\bibnamefont {Lisenfeld}},\ }\href
  {https://doi.org/10.1038/s41534-020-00359-x} {\bibfield  {journal} {\bibinfo
  {journal} {npj Quantum Inf.}\ }\textbf {\bibinfo {volume} {7}},\ \bibinfo
  {pages} {1} (\bibinfo {year} {2021})}\BibitemShut {NoStop}%
\bibitem [{\citenamefont {Sarabi}\ \emph {et~al.}(2015)\citenamefont {Sarabi},
  \citenamefont {Ramanayaka}, \citenamefont {Burin}, \citenamefont
  {Wellstood},\ and\ \citenamefont {Osborn}}]{Sarabi2015}%
  \BibitemOpen
  \bibfield  {author} {\bibinfo {author} {\bibfnamefont {B.}~\bibnamefont
  {Sarabi}}, \bibinfo {author} {\bibfnamefont {A.~N.}\ \bibnamefont
  {Ramanayaka}}, \bibinfo {author} {\bibfnamefont {A.~L.}\ \bibnamefont
  {Burin}}, \bibinfo {author} {\bibfnamefont {F.~C.}\ \bibnamefont
  {Wellstood}},\ and\ \bibinfo {author} {\bibfnamefont {K.~D.}\ \bibnamefont
  {Osborn}},\ }\href {https://doi.org/10.1063/1.4918775} {\bibfield  {journal}
  {\bibinfo  {journal} {Appl. Phys. Lett.}\ }\textbf {\bibinfo {volume}
  {106}},\ \bibinfo {pages} {172601} (\bibinfo {year} {2015})}\BibitemShut
  {NoStop}%
\bibitem [{\citenamefont {Brehm}\ \emph {et~al.}(2017)\citenamefont {Brehm},
  \citenamefont {Bilmes}, \citenamefont {Weiss}, \citenamefont {Ustinov},\ and\
  \citenamefont {Lisenfeld}}]{Brehm2017}%
  \BibitemOpen
  \bibfield  {author} {\bibinfo {author} {\bibfnamefont {J.~D.}\ \bibnamefont
  {Brehm}}, \bibinfo {author} {\bibfnamefont {A.}~\bibnamefont {Bilmes}},
  \bibinfo {author} {\bibfnamefont {G.}~\bibnamefont {Weiss}}, \bibinfo
  {author} {\bibfnamefont {A.~V.}\ \bibnamefont {Ustinov}},\ and\ \bibinfo
  {author} {\bibfnamefont {J.}~\bibnamefont {Lisenfeld}},\ }\href
  {https://doi.org/10.1063/1.5001920} {\bibfield  {journal} {\bibinfo
  {journal} {Appl. Phys. Lett.}\ }\textbf {\bibinfo {volume} {111}},\ \bibinfo
  {pages} {112601} (\bibinfo {year} {2017})}\BibitemShut {NoStop}%
\bibitem [{\citenamefont {Gao}\ \emph {et~al.}(2007)\citenamefont {Gao},
  \citenamefont {Zmuidzinas}, \citenamefont {Mazin}, \citenamefont {LeDuc},\
  and\ \citenamefont {Day}}]{Gao2007}%
  \BibitemOpen
  \bibfield  {author} {\bibinfo {author} {\bibfnamefont {J.}~\bibnamefont
  {Gao}}, \bibinfo {author} {\bibfnamefont {J.}~\bibnamefont {Zmuidzinas}},
  \bibinfo {author} {\bibfnamefont {B.~A.}\ \bibnamefont {Mazin}}, \bibinfo
  {author} {\bibfnamefont {H.~G.}\ \bibnamefont {LeDuc}},\ and\ \bibinfo
  {author} {\bibfnamefont {P.~K.}\ \bibnamefont {Day}},\ }\href
  {https://doi.org/10.1063/1.2711770} {\bibfield  {journal} {\bibinfo
  {journal} {Appl. Phys. Lett.}\ }\textbf {\bibinfo {volume} {90}},\ \bibinfo
  {pages} {102507} (\bibinfo {year} {2007})}\BibitemShut {NoStop}%
\bibitem [{\citenamefont {Pappas}\ \emph {et~al.}(2011)\citenamefont {Pappas},
  \citenamefont {Vissers}, \citenamefont {Wisbey}, \citenamefont {Kline},\ and\
  \citenamefont {Gao}}]{Pappas2011}%
  \BibitemOpen
  \bibfield  {author} {\bibinfo {author} {\bibfnamefont {D.~P.}\ \bibnamefont
  {Pappas}}, \bibinfo {author} {\bibfnamefont {M.~R.}\ \bibnamefont {Vissers}},
  \bibinfo {author} {\bibfnamefont {D.~S.}\ \bibnamefont {Wisbey}}, \bibinfo
  {author} {\bibfnamefont {J.~S.}\ \bibnamefont {Kline}},\ and\ \bibinfo
  {author} {\bibfnamefont {J.}~\bibnamefont {Gao}},\ }\href
  {https://doi.org/10.1109/TASC.2010.2097578} {\bibfield  {journal} {\bibinfo
  {journal} {IEEE Transactions on Applied Superconductivity}\ }\textbf
  {\bibinfo {volume} {21}},\ \bibinfo {pages} {871} (\bibinfo {year}
  {2011})}\BibitemShut {NoStop}%
\bibitem [{\citenamefont {Burnett}\ \emph {et~al.}(2014)\citenamefont
  {Burnett}, \citenamefont {Faoro}, \citenamefont {Wisby}, \citenamefont
  {Gurtovoi}, \citenamefont {Chernykh}, \citenamefont {Mikhailov},
  \citenamefont {Tulin}, \citenamefont {Shaikhaidarov}, \citenamefont
  {Antonov}, \citenamefont {Meeson}, \citenamefont {Tzalenchuk},\ and\
  \citenamefont {Lindström}}]{Burnett2014}%
  \BibitemOpen
  \bibfield  {author} {\bibinfo {author} {\bibfnamefont {J.}~\bibnamefont
  {Burnett}}, \bibinfo {author} {\bibfnamefont {L.}~\bibnamefont {Faoro}},
  \bibinfo {author} {\bibfnamefont {I.}~\bibnamefont {Wisby}}, \bibinfo
  {author} {\bibfnamefont {V.~L.}\ \bibnamefont {Gurtovoi}}, \bibinfo {author}
  {\bibfnamefont {A.~V.}\ \bibnamefont {Chernykh}}, \bibinfo {author}
  {\bibfnamefont {G.~M.}\ \bibnamefont {Mikhailov}}, \bibinfo {author}
  {\bibfnamefont {V.~A.}\ \bibnamefont {Tulin}}, \bibinfo {author}
  {\bibfnamefont {R.}~\bibnamefont {Shaikhaidarov}}, \bibinfo {author}
  {\bibfnamefont {V.}~\bibnamefont {Antonov}}, \bibinfo {author} {\bibfnamefont
  {P.~J.}\ \bibnamefont {Meeson}}, \bibinfo {author} {\bibfnamefont {A.~Y.}\
  \bibnamefont {Tzalenchuk}},\ and\ \bibinfo {author} {\bibfnamefont
  {T.}~\bibnamefont {Lindström}},\ }\href {https://doi.org/10.1038/ncomms5119}
  {\bibfield  {journal} {\bibinfo  {journal} {Nat. Commun.}\ }\textbf {\bibinfo
  {volume} {5}},\ \bibinfo {pages} {4119} (\bibinfo {year} {2014})}\BibitemShut
  {NoStop}%
\bibitem [{\citenamefont {Goetz}\ \emph {et~al.}(2016)\citenamefont {Goetz},
  \citenamefont {Deppe}, \citenamefont {Haeberlein}, \citenamefont {Wulschner},
  \citenamefont {Zollitsch}, \citenamefont {Meier}, \citenamefont {Fischer},
  \citenamefont {Eder}, \citenamefont {Xie}, \citenamefont {Fedorov},
  \citenamefont {Menzel}, \citenamefont {Marx},\ and\ \citenamefont
  {Gross}}]{Goetz2016}%
  \BibitemOpen
  \bibfield  {author} {\bibinfo {author} {\bibfnamefont {J.}~\bibnamefont
  {Goetz}}, \bibinfo {author} {\bibfnamefont {F.}~\bibnamefont {Deppe}},
  \bibinfo {author} {\bibfnamefont {M.}~\bibnamefont {Haeberlein}}, \bibinfo
  {author} {\bibfnamefont {F.}~\bibnamefont {Wulschner}}, \bibinfo {author}
  {\bibfnamefont {C.~W.}\ \bibnamefont {Zollitsch}}, \bibinfo {author}
  {\bibfnamefont {S.}~\bibnamefont {Meier}}, \bibinfo {author} {\bibfnamefont
  {M.}~\bibnamefont {Fischer}}, \bibinfo {author} {\bibfnamefont
  {P.}~\bibnamefont {Eder}}, \bibinfo {author} {\bibfnamefont {E.}~\bibnamefont
  {Xie}}, \bibinfo {author} {\bibfnamefont {K.~G.}\ \bibnamefont {Fedorov}},
  \bibinfo {author} {\bibfnamefont {E.~P.}\ \bibnamefont {Menzel}}, \bibinfo
  {author} {\bibfnamefont {A.}~\bibnamefont {Marx}},\ and\ \bibinfo {author}
  {\bibfnamefont {R.}~\bibnamefont {Gross}},\ }\href
  {https://doi.org/10.1063/1.4939299} {\bibfield  {journal} {\bibinfo
  {journal} {J. Appl. Phys.}\ }\textbf {\bibinfo {volume} {119}},\ \bibinfo
  {pages} {015304} (\bibinfo {year} {2016})}\BibitemShut {NoStop}%
\bibitem [{\citenamefont {Niepce}\ \emph {et~al.}(2020)\citenamefont {Niepce},
  \citenamefont {Burnett}, \citenamefont {Latorre},\ and\ \citenamefont
  {Bylander}}]{Niepce2020}%
  \BibitemOpen
  \bibfield  {author} {\bibinfo {author} {\bibfnamefont {D.}~\bibnamefont
  {Niepce}}, \bibinfo {author} {\bibfnamefont {J.~J.}\ \bibnamefont {Burnett}},
  \bibinfo {author} {\bibfnamefont {M.~G.}\ \bibnamefont {Latorre}},\ and\
  \bibinfo {author} {\bibfnamefont {J.}~\bibnamefont {Bylander}},\ }\href
  {https://doi.org/10.1088/1361-6668/ab6179} {\bibfield  {journal} {\bibinfo
  {journal} {Supercond. Sci. Technol.}\ }\textbf {\bibinfo {volume} {33}},\
  \bibinfo {pages} {025013} (\bibinfo {year} {2020})}\BibitemShut {NoStop}%
\bibitem [{\citenamefont {Deutscher}\ \emph
  {et~al.}(1973{\natexlab{a}})\citenamefont {Deutscher}, \citenamefont
  {Fenichel}, \citenamefont {Gershenson}, \citenamefont {Grünbaum},\ and\
  \citenamefont {Ovadyahu}}]{Deutscher1973}%
  \BibitemOpen
  \bibfield  {author} {\bibinfo {author} {\bibfnamefont {G.}~\bibnamefont
  {Deutscher}}, \bibinfo {author} {\bibfnamefont {H.}~\bibnamefont {Fenichel}},
  \bibinfo {author} {\bibfnamefont {M.}~\bibnamefont {Gershenson}}, \bibinfo
  {author} {\bibfnamefont {E.}~\bibnamefont {Grünbaum}},\ and\ \bibinfo
  {author} {\bibfnamefont {Z.}~\bibnamefont {Ovadyahu}},\ }\href
  {https://doi.org/10.1007/BF00655256} {\bibfield  {journal} {\bibinfo
  {journal} {J. Low Temp. Phys.}\ }\textbf {\bibinfo {volume} {10}},\ \bibinfo
  {pages} {231} (\bibinfo {year} {1973}{\natexlab{a}})}\BibitemShut {NoStop}%
\bibitem [{\citenamefont {Wildermuth}\ \emph {et~al.}(2022)\citenamefont
  {Wildermuth}, \citenamefont {Powalla}, \citenamefont {Voss}, \citenamefont
  {Schön}, \citenamefont {Schneider}, \citenamefont {Fistul}, \citenamefont
  {Rotzinger},\ and\ \citenamefont {Ustinov}}]{Wildermuth2022}%
  \BibitemOpen
  \bibfield  {author} {\bibinfo {author} {\bibfnamefont {M.}~\bibnamefont
  {Wildermuth}}, \bibinfo {author} {\bibfnamefont {L.}~\bibnamefont {Powalla}},
  \bibinfo {author} {\bibfnamefont {J.~N.}\ \bibnamefont {Voss}}, \bibinfo
  {author} {\bibfnamefont {Y.}~\bibnamefont {Schön}}, \bibinfo {author}
  {\bibfnamefont {A.}~\bibnamefont {Schneider}}, \bibinfo {author}
  {\bibfnamefont {M.~V.}\ \bibnamefont {Fistul}}, \bibinfo {author}
  {\bibfnamefont {H.}~\bibnamefont {Rotzinger}},\ and\ \bibinfo {author}
  {\bibfnamefont {A.~V.}\ \bibnamefont {Ustinov}},\ }\href
  {https://doi.org/10.1063/5.0082197} {\bibfield  {journal} {\bibinfo
  {journal} {Appl. Phys. Lett.}\ }\textbf {\bibinfo {volume} {120}},\ \bibinfo
  {pages} {112601} (\bibinfo {year} {2022})}\BibitemShut {NoStop}%
\bibitem [{\citenamefont {Lisenfeld}\ \emph {et~al.}(2019)\citenamefont
  {Lisenfeld}, \citenamefont {Bilmes}, \citenamefont {Megrant}, \citenamefont
  {Barends}, \citenamefont {Kelly}, \citenamefont {Klimov}, \citenamefont
  {Weiss}, \citenamefont {Martinis},\ and\ \citenamefont
  {Ustinov}}]{Lisenfeld2019}%
  \BibitemOpen
  \bibfield  {author} {\bibinfo {author} {\bibfnamefont {J.}~\bibnamefont
  {Lisenfeld}}, \bibinfo {author} {\bibfnamefont {A.}~\bibnamefont {Bilmes}},
  \bibinfo {author} {\bibfnamefont {A.}~\bibnamefont {Megrant}}, \bibinfo
  {author} {\bibfnamefont {R.}~\bibnamefont {Barends}}, \bibinfo {author}
  {\bibfnamefont {J.}~\bibnamefont {Kelly}}, \bibinfo {author} {\bibfnamefont
  {P.}~\bibnamefont {Klimov}}, \bibinfo {author} {\bibfnamefont
  {G.}~\bibnamefont {Weiss}}, \bibinfo {author} {\bibfnamefont {J.~M.}\
  \bibnamefont {Martinis}},\ and\ \bibinfo {author} {\bibfnamefont {A.~V.}\
  \bibnamefont {Ustinov}},\ }\href {https://doi.org/10.1038/s41534-019-0224-1}
  {\bibfield  {journal} {\bibinfo  {journal} {npj Quantum Inf.}\ }\textbf
  {\bibinfo {volume} {5}},\ \bibinfo {pages} {1} (\bibinfo {year}
  {2019})}\BibitemShut {NoStop}%
\bibitem [{\citenamefont {Schneider}(2020)}]{Schneider2020}%
  \BibitemOpen
  \bibfield  {author} {\bibinfo {author} {\bibfnamefont {A.}~\bibnamefont
  {Schneider}},\ }\emph {\bibinfo {title} {Quantum {Sensing} {Experiments} with
  {Superconducting} {Qubits}}},\ \href {https://doi.org/10.5445/KSP/1000118743}
  {Ph.D. thesis},\ \bibinfo  {school} {Karlsruhe Institute of Technology}
  (\bibinfo {year} {2020})\BibitemShut {NoStop}%
\bibitem [{\citenamefont {Shalibo}\ \emph {et~al.}(2010)\citenamefont
  {Shalibo}, \citenamefont {Rofe}, \citenamefont {Shwa}, \citenamefont
  {Zeides}, \citenamefont {Neeley}, \citenamefont {Martinis},\ and\
  \citenamefont {Katz}}]{Shalibo2010}%
  \BibitemOpen
  \bibfield  {author} {\bibinfo {author} {\bibfnamefont {Y.}~\bibnamefont
  {Shalibo}}, \bibinfo {author} {\bibfnamefont {Y.}~\bibnamefont {Rofe}},
  \bibinfo {author} {\bibfnamefont {D.}~\bibnamefont {Shwa}}, \bibinfo {author}
  {\bibfnamefont {F.}~\bibnamefont {Zeides}}, \bibinfo {author} {\bibfnamefont
  {M.}~\bibnamefont {Neeley}}, \bibinfo {author} {\bibfnamefont {J.~M.}\
  \bibnamefont {Martinis}},\ and\ \bibinfo {author} {\bibfnamefont
  {N.}~\bibnamefont {Katz}},\ }\href
  {https://doi.org/10.1103/PhysRevLett.105.177001} {\bibfield  {journal}
  {\bibinfo  {journal} {Phys. Rev. Lett.}\ }\textbf {\bibinfo {volume} {105}},\
  \bibinfo {pages} {177001} (\bibinfo {year} {2010})}\BibitemShut {NoStop}%
\bibitem [{\citenamefont {Faoro}\ and\ \citenamefont
  {Ioffe}(2015)}]{Faoro2015}%
  \BibitemOpen
  \bibfield  {author} {\bibinfo {author} {\bibfnamefont {L.}~\bibnamefont
  {Faoro}}\ and\ \bibinfo {author} {\bibfnamefont {L.~B.}\ \bibnamefont
  {Ioffe}},\ }\href
  {https://doi.org/https://doi.org/10.1103/PhysRevB.91.014201} {\bibfield
  {journal} {\bibinfo  {journal} {Phys. Rev. B}\ }\textbf {\bibinfo {volume}
  {91}},\ \bibinfo {pages} {014201} (\bibinfo {year} {2015})}\BibitemShut
  {NoStop}%
\bibitem [{\citenamefont {Müller}\ \emph {et~al.}(2015)\citenamefont
  {Müller}, \citenamefont {Lisenfeld}, \citenamefont {Shnirman},\ and\
  \citenamefont {Poletto}}]{Muller2015}%
  \BibitemOpen
  \bibfield  {author} {\bibinfo {author} {\bibfnamefont {C.}~\bibnamefont
  {Müller}}, \bibinfo {author} {\bibfnamefont {J.}~\bibnamefont {Lisenfeld}},
  \bibinfo {author} {\bibfnamefont {A.}~\bibnamefont {Shnirman}},\ and\
  \bibinfo {author} {\bibfnamefont {S.}~\bibnamefont {Poletto}},\ }\href
  {https://doi.org/10.1103/PhysRevB.92.035442} {\bibfield  {journal} {\bibinfo
  {journal} {Phys. Rev. B}\ }\textbf {\bibinfo {volume} {92}},\ \bibinfo
  {pages} {035442} (\bibinfo {year} {2015})}\BibitemShut {NoStop}%
\bibitem [{\citenamefont {Schlör}\ \emph {et~al.}(2019)\citenamefont
  {Schlör}, \citenamefont {Lisenfeld}, \citenamefont {Müller}, \citenamefont
  {Bilmes}, \citenamefont {Schneider}, \citenamefont {Pappas}, \citenamefont
  {Ustinov},\ and\ \citenamefont {Weides}}]{Schlor2019}%
  \BibitemOpen
  \bibfield  {author} {\bibinfo {author} {\bibfnamefont {S.}~\bibnamefont
  {Schlör}}, \bibinfo {author} {\bibfnamefont {J.}~\bibnamefont {Lisenfeld}},
  \bibinfo {author} {\bibfnamefont {C.}~\bibnamefont {Müller}}, \bibinfo
  {author} {\bibfnamefont {A.}~\bibnamefont {Bilmes}}, \bibinfo {author}
  {\bibfnamefont {A.}~\bibnamefont {Schneider}}, \bibinfo {author}
  {\bibfnamefont {D.~P.}\ \bibnamefont {Pappas}}, \bibinfo {author}
  {\bibfnamefont {A.~V.}\ \bibnamefont {Ustinov}},\ and\ \bibinfo {author}
  {\bibfnamefont {M.}~\bibnamefont {Weides}},\ }\href
  {https://doi.org/10.1103/PhysRevLett.123.190502} {\bibfield  {journal}
  {\bibinfo  {journal} {Phys. Rev. Lett.}\ }\textbf {\bibinfo {volume} {123}},\
  \bibinfo {pages} {190502} (\bibinfo {year} {2019})}\BibitemShut {NoStop}%
\bibitem [{\citenamefont {Faoro}\ and\ \citenamefont
  {Ioffe}(2012)}]{Faoro2012}%
  \BibitemOpen
  \bibfield  {author} {\bibinfo {author} {\bibfnamefont {L.}~\bibnamefont
  {Faoro}}\ and\ \bibinfo {author} {\bibfnamefont {L.~B.}\ \bibnamefont
  {Ioffe}},\ }\href {https://doi.org/10.1103/PhysRevLett.109.157005} {\bibfield
   {journal} {\bibinfo  {journal} {Phys. Rev. Lett.}\ }\textbf {\bibinfo
  {volume} {109}},\ \bibinfo {pages} {157005} (\bibinfo {year}
  {2012})}\BibitemShut {NoStop}%
\bibitem [{\citenamefont {Anderson}\ \emph {et~al.}(1972)\citenamefont
  {Anderson}, \citenamefont {Halperin},\ and\ \citenamefont
  {Varma}}]{Anderson1972}%
  \BibitemOpen
  \bibfield  {author} {\bibinfo {author} {\bibfnamefont {P.~w.}\ \bibnamefont
  {Anderson}}, \bibinfo {author} {\bibfnamefont {B.~I.}\ \bibnamefont
  {Halperin}},\ and\ \bibinfo {author} {\bibfnamefont {c.~M.}\ \bibnamefont
  {Varma}},\ }\href {https://doi.org/10.1080/14786437208229210} {\bibfield
  {journal} {\bibinfo  {journal} {J. Appl. Phys.}\ }\textbf {\bibinfo {volume}
  {25}},\ \bibinfo {pages} {1} (\bibinfo {year} {1972})}\BibitemShut {NoStop}%
\bibitem [{\citenamefont {Phillips}(1972)}]{Phillips1972}%
  \BibitemOpen
  \bibfield  {author} {\bibinfo {author} {\bibfnamefont {W.~A.}\ \bibnamefont
  {Phillips}},\ }\href {https://doi.org/10.1007/BF00660072} {\bibfield
  {journal} {\bibinfo  {journal} {J. Low Temp. Phys.}\ }\textbf {\bibinfo
  {volume} {7}},\ \bibinfo {pages} {351} (\bibinfo {year} {1972})}\BibitemShut
  {NoStop}%
\bibitem [{\citenamefont {Khalil}\ \emph {et~al.}(2012)\citenamefont {Khalil},
  \citenamefont {Stoutimore}, \citenamefont {Wellstood},\ and\ \citenamefont
  {Osborn}}]{Khalil2012}%
  \BibitemOpen
  \bibfield  {author} {\bibinfo {author} {\bibfnamefont {M.~S.}\ \bibnamefont
  {Khalil}}, \bibinfo {author} {\bibfnamefont {M.~J.~A.}\ \bibnamefont
  {Stoutimore}}, \bibinfo {author} {\bibfnamefont {F.~C.}\ \bibnamefont
  {Wellstood}},\ and\ \bibinfo {author} {\bibfnamefont {K.~D.}\ \bibnamefont
  {Osborn}},\ }\href {https://doi.org/10.1063/1.3692073} {\bibfield  {journal}
  {\bibinfo  {journal} {J. Appl. Phys.}\ }\textbf {\bibinfo {volume} {111}},\
  \bibinfo {pages} {054510} (\bibinfo {year} {2012})}\BibitemShut {NoStop}%
\bibitem [{\citenamefont {Sarabi}(2014)}]{Sarabi2014}%
  \BibitemOpen
  \bibfield  {author} {\bibinfo {author} {\bibfnamefont {B.}~\bibnamefont
  {Sarabi}},\ }\emph {\bibinfo {title} {Cavity {Quantum} {Electrodynamics} of
  {Nanoscale} {Two}-{Level} {Systems}}},\ \href
  {https://drum.lib.umd.edu/handle/1903/16186} {Ph.D. thesis},\ \bibinfo
  {school} {University of Maryland} (\bibinfo {year} {2014})\BibitemShut
  {NoStop}%
\bibitem [{\citenamefont {Von~Schickfus}\ and\ \citenamefont
  {Hunklinger}(1977)}]{VonSchickfus1977}%
  \BibitemOpen
  \bibfield  {author} {\bibinfo {author} {\bibfnamefont {M.}~\bibnamefont
  {Von~Schickfus}}\ and\ \bibinfo {author} {\bibfnamefont {S.}~\bibnamefont
  {Hunklinger}},\ }\href {https://doi.org/10.1016/0375-9601(77)90558-8}
  {\bibfield  {journal} {\bibinfo  {journal} {Physics Letters A}\ }\textbf
  {\bibinfo {volume} {64}},\ \bibinfo {pages} {144} (\bibinfo {year}
  {1977})}\BibitemShut {NoStop}%
\bibitem [{\citenamefont {Golding}\ \emph {et~al.}(1979)\citenamefont
  {Golding}, \citenamefont {Schickfus}, \citenamefont {Hunklinger},\ and\
  \citenamefont {Dransfeld}}]{Golding1979}%
  \BibitemOpen
  \bibfield  {author} {\bibinfo {author} {\bibfnamefont {B.}~\bibnamefont
  {Golding}}, \bibinfo {author} {\bibfnamefont {M.~v.}\ \bibnamefont
  {Schickfus}}, \bibinfo {author} {\bibfnamefont {S.}~\bibnamefont
  {Hunklinger}},\ and\ \bibinfo {author} {\bibfnamefont {K.}~\bibnamefont
  {Dransfeld}},\ }\href {https://doi.org/10.1103/PhysRevLett.43.1817}
  {\bibfield  {journal} {\bibinfo  {journal} {Phys. Rev. Lett.}\ }\textbf
  {\bibinfo {volume} {43}},\ \bibinfo {pages} {1817} (\bibinfo {year}
  {1979})}\BibitemShut {NoStop}%
\bibitem [{\citenamefont {Martinis}\ \emph {et~al.}(2005)\citenamefont
  {Martinis}, \citenamefont {Cooper}, \citenamefont {McDermott}, \citenamefont
  {Steffen}, \citenamefont {Ansmann}, \citenamefont {Osborn}, \citenamefont
  {Cicak}, \citenamefont {Oh}, \citenamefont {Pappas}, \citenamefont
  {Simmonds},\ and\ \citenamefont {Yu}}]{Martinis2005}%
  \BibitemOpen
  \bibfield  {author} {\bibinfo {author} {\bibfnamefont {J.~M.}\ \bibnamefont
  {Martinis}}, \bibinfo {author} {\bibfnamefont {K.~B.}\ \bibnamefont
  {Cooper}}, \bibinfo {author} {\bibfnamefont {R.}~\bibnamefont {McDermott}},
  \bibinfo {author} {\bibfnamefont {M.}~\bibnamefont {Steffen}}, \bibinfo
  {author} {\bibfnamefont {M.}~\bibnamefont {Ansmann}}, \bibinfo {author}
  {\bibfnamefont {K.~D.}\ \bibnamefont {Osborn}}, \bibinfo {author}
  {\bibfnamefont {K.}~\bibnamefont {Cicak}}, \bibinfo {author} {\bibfnamefont
  {S.}~\bibnamefont {Oh}}, \bibinfo {author} {\bibfnamefont {D.~P.}\
  \bibnamefont {Pappas}}, \bibinfo {author} {\bibfnamefont {R.~W.}\
  \bibnamefont {Simmonds}},\ and\ \bibinfo {author} {\bibfnamefont {C.~C.}\
  \bibnamefont {Yu}},\ }\href {https://doi.org/10.1103/PhysRevLett.95.210503}
  {\bibfield  {journal} {\bibinfo  {journal} {Phys. Rev. Lett.}\ }\textbf
  {\bibinfo {volume} {95}},\ \bibinfo {pages} {210503} (\bibinfo {year}
  {2005})}\BibitemShut {NoStop}%
\bibitem [{\citenamefont {Sarabi}\ \emph {et~al.}(2016)\citenamefont {Sarabi},
  \citenamefont {Ramanayaka}, \citenamefont {Burin}, \citenamefont
  {Wellstood},\ and\ \citenamefont {Osborn}}]{Sarabi2016}%
  \BibitemOpen
  \bibfield  {author} {\bibinfo {author} {\bibfnamefont {B.}~\bibnamefont
  {Sarabi}}, \bibinfo {author} {\bibfnamefont {A.}~\bibnamefont {Ramanayaka}},
  \bibinfo {author} {\bibfnamefont {A.}~\bibnamefont {Burin}}, \bibinfo
  {author} {\bibfnamefont {F.}~\bibnamefont {Wellstood}},\ and\ \bibinfo
  {author} {\bibfnamefont {K.}~\bibnamefont {Osborn}},\ }\href
  {https://doi.org/10.1103/PhysRevLett.116.167002} {\bibfield  {journal}
  {\bibinfo  {journal} {Phys. Rev. Lett.}\ }\textbf {\bibinfo {volume} {116}},\
  \bibinfo {pages} {167002} (\bibinfo {year} {2016})}\BibitemShut {NoStop}%
\bibitem [{\citenamefont {Bilmes}\ \emph {et~al.}(2020)\citenamefont {Bilmes},
  \citenamefont {Megrant}, \citenamefont {Klimov}, \citenamefont {Weiss},
  \citenamefont {Martinis}, \citenamefont {Ustinov},\ and\ \citenamefont
  {Lisenfeld}}]{Bilmes2020}%
  \BibitemOpen
  \bibfield  {author} {\bibinfo {author} {\bibfnamefont {A.}~\bibnamefont
  {Bilmes}}, \bibinfo {author} {\bibfnamefont {A.}~\bibnamefont {Megrant}},
  \bibinfo {author} {\bibfnamefont {P.}~\bibnamefont {Klimov}}, \bibinfo
  {author} {\bibfnamefont {G.}~\bibnamefont {Weiss}}, \bibinfo {author}
  {\bibfnamefont {J.~M.}\ \bibnamefont {Martinis}}, \bibinfo {author}
  {\bibfnamefont {A.~V.}\ \bibnamefont {Ustinov}},\ and\ \bibinfo {author}
  {\bibfnamefont {J.}~\bibnamefont {Lisenfeld}},\ }\href
  {https://doi.org/10.1038/s41598-020-59749-y} {\bibfield  {journal} {\bibinfo
  {journal} {Sci. Rep.}\ }\textbf {\bibinfo {volume} {10}},\ \bibinfo {pages}
  {3090} (\bibinfo {year} {2020})}\BibitemShut {NoStop}%
\bibitem [{\citenamefont {Yu}\ \emph {et~al.}(2022)\citenamefont {Yu},
  \citenamefont {Matityahu}, \citenamefont {Rosen}, \citenamefont {Hung},
  \citenamefont {Maksymov}, \citenamefont {Burin}, \citenamefont {Schechter},\
  and\ \citenamefont {Osborn}}]{Yu2022}%
  \BibitemOpen
  \bibfield  {author} {\bibinfo {author} {\bibfnamefont {L.}~\bibnamefont
  {Yu}}, \bibinfo {author} {\bibfnamefont {S.}~\bibnamefont {Matityahu}},
  \bibinfo {author} {\bibfnamefont {Y.~J.}\ \bibnamefont {Rosen}}, \bibinfo
  {author} {\bibfnamefont {C.-C.}\ \bibnamefont {Hung}}, \bibinfo {author}
  {\bibfnamefont {A.}~\bibnamefont {Maksymov}}, \bibinfo {author}
  {\bibfnamefont {A.~L.}\ \bibnamefont {Burin}}, \bibinfo {author}
  {\bibfnamefont {M.}~\bibnamefont {Schechter}},\ and\ \bibinfo {author}
  {\bibfnamefont {K.~D.}\ \bibnamefont {Osborn}},\ }\href
  {https://doi.org/10.1038/s41598-022-21256-7} {\bibfield  {journal} {\bibinfo
  {journal} {Sci. Rep.}\ }\textbf {\bibinfo {volume} {12}},\ \bibinfo {pages}
  {16960} (\bibinfo {year} {2022})}\BibitemShut {NoStop}%
\bibitem [{\citenamefont {Schechter}\ and\ \citenamefont
  {Stamp}(2013)}]{Schechter2013}%
  \BibitemOpen
  \bibfield  {author} {\bibinfo {author} {\bibfnamefont {M.}~\bibnamefont
  {Schechter}}\ and\ \bibinfo {author} {\bibfnamefont {P.~C.~E.}\ \bibnamefont
  {Stamp}},\ }\href {https://doi.org/10.1103/PhysRevB.88.174202} {\bibfield
  {journal} {\bibinfo  {journal} {Phys. Rev. B}\ }\textbf {\bibinfo {volume}
  {88}},\ \bibinfo {pages} {174202} (\bibinfo {year} {2013})}\BibitemShut
  {NoStop}%
\bibitem [{\citenamefont {Efros}\ and\ \citenamefont
  {Shklovskii}(1975)}]{Efros1975}%
  \BibitemOpen
  \bibfield  {author} {\bibinfo {author} {\bibfnamefont {A.~L.}\ \bibnamefont
  {Efros}}\ and\ \bibinfo {author} {\bibfnamefont {B.~I.}\ \bibnamefont
  {Shklovskii}},\ }\href {https://doi.org/10.1088/0022-3719/8/4/003} {\bibfield
   {journal} {\bibinfo  {journal} {J. Phys. C: Solid State Phys.}\ }\textbf
  {\bibinfo {volume} {8}},\ \bibinfo {pages} {L49} (\bibinfo {year}
  {1975})}\BibitemShut {NoStop}%
\bibitem [{\citenamefont {Churkin}\ \emph {et~al.}(2014)\citenamefont
  {Churkin}, \citenamefont {Barash},\ and\ \citenamefont
  {Schechter}}]{Churkin2014}%
  \BibitemOpen
  \bibfield  {author} {\bibinfo {author} {\bibfnamefont {A.}~\bibnamefont
  {Churkin}}, \bibinfo {author} {\bibfnamefont {D.}~\bibnamefont {Barash}},\
  and\ \bibinfo {author} {\bibfnamefont {M.}~\bibnamefont {Schechter}},\ }\href
  {https://doi.org/10.1103/PhysRevB.89.104202} {\bibfield  {journal} {\bibinfo
  {journal} {Phys. Rev. B}\ }\textbf {\bibinfo {volume} {89}},\ \bibinfo
  {pages} {104202} (\bibinfo {year} {2014})}\BibitemShut {NoStop}%
\bibitem [{\citenamefont {Sidorova}\ \emph {et~al.}(2022)\citenamefont
  {Sidorova}, \citenamefont {Semenov}, \citenamefont {Hübers}, \citenamefont
  {Gyger},\ and\ \citenamefont {Steinhauer}}]{Sidorova2022}%
  \BibitemOpen
  \bibfield  {author} {\bibinfo {author} {\bibfnamefont {M.}~\bibnamefont
  {Sidorova}}, \bibinfo {author} {\bibfnamefont {A.~D.}\ \bibnamefont
  {Semenov}}, \bibinfo {author} {\bibfnamefont {H.~W.}\ \bibnamefont
  {Hübers}}, \bibinfo {author} {\bibfnamefont {S.}~\bibnamefont {Gyger}},\
  and\ \bibinfo {author} {\bibfnamefont {S.}~\bibnamefont {Steinhauer}},\
  }\href {https://doi.org/10.1088/1361-6668/AC8454} {\bibfield  {journal}
  {\bibinfo  {journal} {Supercond. Sci. Technol.}\ }\textbf {\bibinfo {volume}
  {35}},\ \bibinfo {pages} {105005} (\bibinfo {year} {2022})}\BibitemShut
  {NoStop}%
\bibitem [{\citenamefont {Lubchenko}\ and\ \citenamefont
  {Wolynes}(2001)}]{Lubchenko2001}%
  \BibitemOpen
  \bibfield  {author} {\bibinfo {author} {\bibfnamefont {V.}~\bibnamefont
  {Lubchenko}}\ and\ \bibinfo {author} {\bibfnamefont {P.~G.}\ \bibnamefont
  {Wolynes}},\ }\href {https://doi.org/10.1103/PhysRevLett.87.195901}
  {\bibfield  {journal} {\bibinfo  {journal} {Phys. Rev. Lett.}\ }\textbf
  {\bibinfo {volume} {87}},\ \bibinfo {pages} {195901} (\bibinfo {year}
  {2001})}\BibitemShut {NoStop}%
\bibitem [{\citenamefont {Levy-Bertrand}\ \emph {et~al.}(2019)\citenamefont
  {Levy-Bertrand}, \citenamefont {Klein}, \citenamefont {Grenet}, \citenamefont
  {Dupré}, \citenamefont {Benoît}, \citenamefont {Bideaud}, \citenamefont
  {Bourrion}, \citenamefont {Calvo}, \citenamefont {Catalano}, \citenamefont
  {Gomez}, \citenamefont {Goupy}, \citenamefont {Grünhaupt}, \citenamefont
  {Luepke}, \citenamefont {Maleeva}, \citenamefont {Valenti}, \citenamefont
  {Pop},\ and\ \citenamefont {Monfardini}}]{Bertrand2019}%
  \BibitemOpen
  \bibfield  {author} {\bibinfo {author} {\bibfnamefont {F.}~\bibnamefont
  {Levy-Bertrand}}, \bibinfo {author} {\bibfnamefont {T.}~\bibnamefont
  {Klein}}, \bibinfo {author} {\bibfnamefont {T.}~\bibnamefont {Grenet}},
  \bibinfo {author} {\bibfnamefont {O.}~\bibnamefont {Dupré}}, \bibinfo
  {author} {\bibfnamefont {A.}~\bibnamefont {Benoît}}, \bibinfo {author}
  {\bibfnamefont {A.}~\bibnamefont {Bideaud}}, \bibinfo {author} {\bibfnamefont
  {O.}~\bibnamefont {Bourrion}}, \bibinfo {author} {\bibfnamefont
  {M.}~\bibnamefont {Calvo}}, \bibinfo {author} {\bibfnamefont
  {A.}~\bibnamefont {Catalano}}, \bibinfo {author} {\bibfnamefont
  {A.}~\bibnamefont {Gomez}}, \bibinfo {author} {\bibfnamefont
  {J.}~\bibnamefont {Goupy}}, \bibinfo {author} {\bibfnamefont
  {L.}~\bibnamefont {Grünhaupt}}, \bibinfo {author} {\bibfnamefont {U.~v.}\
  \bibnamefont {Luepke}}, \bibinfo {author} {\bibfnamefont {N.}~\bibnamefont
  {Maleeva}}, \bibinfo {author} {\bibfnamefont {F.}~\bibnamefont {Valenti}},
  \bibinfo {author} {\bibfnamefont {I.~M.}\ \bibnamefont {Pop}},\ and\ \bibinfo
  {author} {\bibfnamefont {A.}~\bibnamefont {Monfardini}},\ }\href
  {https://doi.org/10.1103/PhysRevB.99.094506} {\bibfield  {journal} {\bibinfo
  {journal} {Phys. Rev. B}\ }\textbf {\bibinfo {volume} {99}},\ \bibinfo
  {pages} {094506} (\bibinfo {year} {2019})}\BibitemShut {NoStop}%
\bibitem [{\citenamefont {Kristen}\ \emph {et~al.}(2023)\citenamefont
  {Kristen}, \citenamefont {Voss}, \citenamefont {Wildermuth}, \citenamefont
  {Rotzinger},\ and\ \citenamefont {Ustinov}}]{Kristen2023}%
  \BibitemOpen
  \bibfield  {author} {\bibinfo {author} {\bibfnamefont {M.}~\bibnamefont
  {Kristen}}, \bibinfo {author} {\bibfnamefont {J.~N.}\ \bibnamefont {Voss}},
  \bibinfo {author} {\bibfnamefont {M.}~\bibnamefont {Wildermuth}}, \bibinfo
  {author} {\bibfnamefont {H.}~\bibnamefont {Rotzinger}},\ and\ \bibinfo
  {author} {\bibfnamefont {A.~V.}\ \bibnamefont {Ustinov}},\ }\href
  {https://doi.org/10.1063/5.0147430} {\bibfield  {journal} {\bibinfo
  {journal} {Appl. Phys. Lett.}\ }\textbf {\bibinfo {volume} {122}},\ \bibinfo
  {pages} {202602} (\bibinfo {year} {2023})}\BibitemShut {NoStop}%
\bibitem [{\citenamefont {de~Graaf}\ \emph {et~al.}(2020)\citenamefont
  {de~Graaf}, \citenamefont {Faoro}, \citenamefont {Ioffe}, \citenamefont
  {Mahashabde}, \citenamefont {Burnett}, \citenamefont {Lindström},
  \citenamefont {Kubatkin}, \citenamefont {Danilov},\ and\ \citenamefont
  {Tzalenchuk}}]{deGraaf2020}%
  \BibitemOpen
  \bibfield  {author} {\bibinfo {author} {\bibfnamefont {S.~E.}\ \bibnamefont
  {de~Graaf}}, \bibinfo {author} {\bibfnamefont {L.}~\bibnamefont {Faoro}},
  \bibinfo {author} {\bibfnamefont {L.~B.}\ \bibnamefont {Ioffe}}, \bibinfo
  {author} {\bibfnamefont {S.}~\bibnamefont {Mahashabde}}, \bibinfo {author}
  {\bibfnamefont {J.~J.}\ \bibnamefont {Burnett}}, \bibinfo {author}
  {\bibfnamefont {T.}~\bibnamefont {Lindström}}, \bibinfo {author}
  {\bibfnamefont {S.~E.}\ \bibnamefont {Kubatkin}}, \bibinfo {author}
  {\bibfnamefont {A.~V.}\ \bibnamefont {Danilov}},\ and\ \bibinfo {author}
  {\bibfnamefont {A.~Y.}\ \bibnamefont {Tzalenchuk}},\ }\href
  {https://doi.org/10.1126/sciadv.abc5055} {\bibfield  {journal} {\bibinfo
  {journal} {Sci. Adv.}\ }\textbf {\bibinfo {volume} {6}},\ \bibinfo {pages}
  {eabc5055} (\bibinfo {year} {2020})}\BibitemShut {NoStop}%
\bibitem [{\citenamefont {Bachar}\ \emph {et~al.}(2015)\citenamefont {Bachar},
  \citenamefont {Lerer}, \citenamefont {Levy}, \citenamefont {Hacohen-Gourgy},
  \citenamefont {Almog}, \citenamefont {Saadaoui}, \citenamefont {Salman},
  \citenamefont {Morenzoni},\ and\ \citenamefont {Deutscher}}]{Bachar2015}%
  \BibitemOpen
  \bibfield  {author} {\bibinfo {author} {\bibfnamefont {N.}~\bibnamefont
  {Bachar}}, \bibinfo {author} {\bibfnamefont {S.}~\bibnamefont {Lerer}},
  \bibinfo {author} {\bibfnamefont {A.}~\bibnamefont {Levy}}, \bibinfo {author}
  {\bibfnamefont {S.}~\bibnamefont {Hacohen-Gourgy}}, \bibinfo {author}
  {\bibfnamefont {B.}~\bibnamefont {Almog}}, \bibinfo {author} {\bibfnamefont
  {H.}~\bibnamefont {Saadaoui}}, \bibinfo {author} {\bibfnamefont
  {Z.}~\bibnamefont {Salman}}, \bibinfo {author} {\bibfnamefont
  {E.}~\bibnamefont {Morenzoni}},\ and\ \bibinfo {author} {\bibfnamefont
  {G.}~\bibnamefont {Deutscher}},\ }\href
  {https://doi.org/10.1103/PHYSREVB.91.041123/FIGURES/3/MEDIUM} {\bibfield
  {journal} {\bibinfo  {journal} {Phys. Rev. B}\ }\textbf {\bibinfo {volume}
  {91}},\ \bibinfo {pages} {041123} (\bibinfo {year} {2015})}\BibitemShut
  {NoStop}%
\bibitem [{\citenamefont {Yang}\ \emph {et~al.}(2020)\citenamefont {Yang},
  \citenamefont {Gozlinski}, \citenamefont {Storbeck}, \citenamefont
  {Grünhaupt}, \citenamefont {Pop},\ and\ \citenamefont
  {Wulfhekel}}]{Yang2020}%
  \BibitemOpen
  \bibfield  {author} {\bibinfo {author} {\bibfnamefont {F.}~\bibnamefont
  {Yang}}, \bibinfo {author} {\bibfnamefont {T.}~\bibnamefont {Gozlinski}},
  \bibinfo {author} {\bibfnamefont {T.}~\bibnamefont {Storbeck}}, \bibinfo
  {author} {\bibfnamefont {L.}~\bibnamefont {Grünhaupt}}, \bibinfo {author}
  {\bibfnamefont {I.~M.}\ \bibnamefont {Pop}},\ and\ \bibinfo {author}
  {\bibfnamefont {W.}~\bibnamefont {Wulfhekel}},\ }\href
  {https://doi.org/10.1103/PhysRevB.102.104502} {\bibfield  {journal} {\bibinfo
   {journal} {Phys. Rev. B}\ }\textbf {\bibinfo {volume} {102}},\ \bibinfo
  {pages} {104502} (\bibinfo {year} {2020})}\BibitemShut {NoStop}%
\bibitem [{\citenamefont {Efetov}(1980)}]{Efetov1980}%
  \BibitemOpen
  \bibfield  {author} {\bibinfo {author} {\bibfnamefont {K.~B.}\ \bibnamefont
  {Efetov}},\ }\href {https://www.osti.gov/biblio/7061039} {\bibfield
  {journal} {\bibinfo  {journal} {Sov. Phys. JETP}\ }\textbf {\bibinfo {volume}
  {51}} (\bibinfo {year} {1980})}\BibitemShut {NoStop}%
\bibitem [{\citenamefont {Pracht}\ \emph {et~al.}(2016)\citenamefont {Pracht},
  \citenamefont {Bachar}, \citenamefont {Benfatto}, \citenamefont {Deutscher},
  \citenamefont {Farber}, \citenamefont {Dressel},\ and\ \citenamefont
  {Scheffler}}]{Pracht2016}%
  \BibitemOpen
  \bibfield  {author} {\bibinfo {author} {\bibfnamefont {U.~S.}\ \bibnamefont
  {Pracht}}, \bibinfo {author} {\bibfnamefont {N.}~\bibnamefont {Bachar}},
  \bibinfo {author} {\bibfnamefont {L.}~\bibnamefont {Benfatto}}, \bibinfo
  {author} {\bibfnamefont {G.}~\bibnamefont {Deutscher}}, \bibinfo {author}
  {\bibfnamefont {E.}~\bibnamefont {Farber}}, \bibinfo {author} {\bibfnamefont
  {M.}~\bibnamefont {Dressel}},\ and\ \bibinfo {author} {\bibfnamefont
  {M.}~\bibnamefont {Scheffler}},\ }\href
  {https://doi.org/10.1103/PhysRevB.93.100503} {\bibfield  {journal} {\bibinfo
  {journal} {Phys. Rev. B}\ }\textbf {\bibinfo {volume} {93}},\ \bibinfo
  {pages} {100503} (\bibinfo {year} {2016})}\BibitemShut {NoStop}%
\bibitem [{\citenamefont {Humbert}\ \emph {et~al.}(2021)\citenamefont
  {Humbert}, \citenamefont {Ortuño}, \citenamefont {Somoza}, \citenamefont
  {Bergé}, \citenamefont {Dumoulin},\ and\ \citenamefont
  {Marrache-Kikuchi}}]{Humbert2021}%
  \BibitemOpen
  \bibfield  {author} {\bibinfo {author} {\bibfnamefont {V.}~\bibnamefont
  {Humbert}}, \bibinfo {author} {\bibfnamefont {M.}~\bibnamefont {Ortuño}},
  \bibinfo {author} {\bibfnamefont {A.~M.}\ \bibnamefont {Somoza}}, \bibinfo
  {author} {\bibfnamefont {L.}~\bibnamefont {Bergé}}, \bibinfo {author}
  {\bibfnamefont {L.}~\bibnamefont {Dumoulin}},\ and\ \bibinfo {author}
  {\bibfnamefont {C.~A.}\ \bibnamefont {Marrache-Kikuchi}},\ }\href
  {https://doi.org/10.1038/s41467-021-26911-7} {\bibfield  {journal} {\bibinfo
  {journal} {Nat. Commun.}\ }\textbf {\bibinfo {volume} {12}},\ \bibinfo
  {pages} {1} (\bibinfo {year} {2021})}\BibitemShut {NoStop}%
\bibitem [{\citenamefont {Koch}\ \emph {et~al.}(2007)\citenamefont {Koch},
  \citenamefont {DiVincenzo},\ and\ \citenamefont {Clarke}}]{Koch2007a}%
  \BibitemOpen
  \bibfield  {author} {\bibinfo {author} {\bibfnamefont {R.~H.}\ \bibnamefont
  {Koch}}, \bibinfo {author} {\bibfnamefont {D.~P.}\ \bibnamefont
  {DiVincenzo}},\ and\ \bibinfo {author} {\bibfnamefont {J.}~\bibnamefont
  {Clarke}},\ }\href {https://doi.org/10.1103/PhysRevLett.98.267003} {\bibfield
   {journal} {\bibinfo  {journal} {Phys. Rev. Lett.}\ }\textbf {\bibinfo
  {volume} {98}},\ \bibinfo {pages} {267003} (\bibinfo {year}
  {2007})}\BibitemShut {NoStop}%
\bibitem [{\citenamefont {Lutchyn}\ \emph {et~al.}(2008)\citenamefont
  {Lutchyn}, \citenamefont {Cywinski}, \citenamefont {Nave},\ and\
  \citenamefont {Das~Sarma}}]{Lutchyn2008}%
  \BibitemOpen
  \bibfield  {author} {\bibinfo {author} {\bibfnamefont {R.~M.}\ \bibnamefont
  {Lutchyn}}, \bibinfo {author} {\bibfnamefont {L.}~\bibnamefont {Cywinski}},
  \bibinfo {author} {\bibfnamefont {C.~P.}\ \bibnamefont {Nave}},\ and\
  \bibinfo {author} {\bibfnamefont {S.}~\bibnamefont {Das~Sarma}},\ }\href
  {https://doi.org/10.1103/PhysRevB.78.024508} {\bibfield  {journal} {\bibinfo
  {journal} {Phys. Rev. B}\ }\textbf {\bibinfo {volume} {78}},\ \bibinfo
  {pages} {024508} (\bibinfo {year} {2008})}\BibitemShut {NoStop}%
\bibitem [{\citenamefont {Burnett}\ \emph {et~al.}(2016)\citenamefont
  {Burnett}, \citenamefont {Faoro},\ and\ \citenamefont
  {Lindström}}]{Burnett2016}%
  \BibitemOpen
  \bibfield  {author} {\bibinfo {author} {\bibfnamefont {J.}~\bibnamefont
  {Burnett}}, \bibinfo {author} {\bibfnamefont {L.}~\bibnamefont {Faoro}},\
  and\ \bibinfo {author} {\bibfnamefont {T.}~\bibnamefont {Lindström}},\
  }\href {https://doi.org/10.1088/0953-2048/29/4/044008} {\bibfield  {journal}
  {\bibinfo  {journal} {Supercond. Sci. Technol.}\ }\textbf {\bibinfo {volume}
  {29}},\ \bibinfo {pages} {044008} (\bibinfo {year} {2016})}\BibitemShut
  {NoStop}%
\bibitem [{\citenamefont {Agarwal}\ \emph {et~al.}(2013)\citenamefont
  {Agarwal}, \citenamefont {Martin}, \citenamefont {Lukin},\ and\ \citenamefont
  {Demler}}]{Agarwal2013}%
  \BibitemOpen
  \bibfield  {author} {\bibinfo {author} {\bibfnamefont {K.}~\bibnamefont
  {Agarwal}}, \bibinfo {author} {\bibfnamefont {I.}~\bibnamefont {Martin}},
  \bibinfo {author} {\bibfnamefont {M.~D.}\ \bibnamefont {Lukin}},\ and\
  \bibinfo {author} {\bibfnamefont {E.}~\bibnamefont {Demler}},\ }\href
  {https://doi.org/10.1103/PhysRevB.87.144201} {\bibfield  {journal} {\bibinfo
  {journal} {Phys. Rev. B}\ }\textbf {\bibinfo {volume} {87}},\ \bibinfo
  {pages} {144201} (\bibinfo {year} {2013})}\BibitemShut {NoStop}%
\bibitem [{\citenamefont {Chakravarty}\ \emph {et~al.}(1987)\citenamefont
  {Chakravarty}, \citenamefont {Kivelson}, \citenamefont {Zimanyi},\ and\
  \citenamefont {Halperin}}]{Chakravarty1987}%
  \BibitemOpen
  \bibfield  {author} {\bibinfo {author} {\bibfnamefont {S.}~\bibnamefont
  {Chakravarty}}, \bibinfo {author} {\bibfnamefont {S.}~\bibnamefont
  {Kivelson}}, \bibinfo {author} {\bibfnamefont {G.~T.}\ \bibnamefont
  {Zimanyi}},\ and\ \bibinfo {author} {\bibfnamefont {B.~I.}\ \bibnamefont
  {Halperin}},\ }\href {https://doi.org/10.1103/PhysRevB.35.7256} {\bibfield
  {journal} {\bibinfo  {journal} {Phys. Rev. B}\ }\textbf {\bibinfo {volume}
  {35}},\ \bibinfo {pages} {7256} (\bibinfo {year} {1987})}\BibitemShut
  {NoStop}%
\bibitem [{\citenamefont {Raychaudhuri}\ and\ \citenamefont
  {Dutta}(2021)}]{Raychaudhuri2021}%
  \BibitemOpen
  \bibfield  {author} {\bibinfo {author} {\bibfnamefont {P.}~\bibnamefont
  {Raychaudhuri}}\ and\ \bibinfo {author} {\bibfnamefont {S.}~\bibnamefont
  {Dutta}},\ }\href {https://doi.org/10.1088/1361-648X/ac360b} {\bibfield
  {journal} {\bibinfo  {journal} {J. Phys.: Condens. Matter}\ }\textbf
  {\bibinfo {volume} {34}},\ \bibinfo {pages} {083001} (\bibinfo {year}
  {2021})}\BibitemShut {NoStop}%
\bibitem [{\citenamefont {Khvalyuk}\ \emph {et~al.}()\citenamefont {Khvalyuk},
  \citenamefont {Charpentier}, \citenamefont {Roch}, \citenamefont {Sacépé},\
  and\ \citenamefont {Feigel'man}}]{Khvalyuk2023a}%
  \BibitemOpen
  \bibfield  {author} {\bibinfo {author} {\bibfnamefont {A.~V.}\ \bibnamefont
  {Khvalyuk}}, \bibinfo {author} {\bibfnamefont {T.}~\bibnamefont
  {Charpentier}}, \bibinfo {author} {\bibfnamefont {N.}~\bibnamefont {Roch}},
  \bibinfo {author} {\bibfnamefont {B.}~\bibnamefont {Sacépé}},\ and\
  \bibinfo {author} {\bibfnamefont {M.~V.}\ \bibnamefont {Feigel'man}},\ }\href
  {http://arxiv.org/abs/2311.15126} {\bibinfo  {journal} {arXiv:2311.15126}\
  }\BibitemShut {NoStop}%
\bibitem [{\citenamefont {Pracht}\ \emph {et~al.}(2017)\citenamefont {Pracht},
  \citenamefont {Cea}, \citenamefont {Bachar}, \citenamefont {Deutscher},
  \citenamefont {Farber}, \citenamefont {Dressel}, \citenamefont {Scheffler},
  \citenamefont {Castellani}, \citenamefont {García-García},\ and\
  \citenamefont {Benfatto}}]{Pracht2017}%
  \BibitemOpen
\bibfield  {journal} {  }\bibfield  {author} {\bibinfo {author} {\bibfnamefont
  {U.~S.}\ \bibnamefont {Pracht}}, \bibinfo {author} {\bibfnamefont
  {T.}~\bibnamefont {Cea}}, \bibinfo {author} {\bibfnamefont {N.}~\bibnamefont
  {Bachar}}, \bibinfo {author} {\bibfnamefont {G.}~\bibnamefont {Deutscher}},
  \bibinfo {author} {\bibfnamefont {E.}~\bibnamefont {Farber}}, \bibinfo
  {author} {\bibfnamefont {M.}~\bibnamefont {Dressel}}, \bibinfo {author}
  {\bibfnamefont {M.}~\bibnamefont {Scheffler}}, \bibinfo {author}
  {\bibfnamefont {C.}~\bibnamefont {Castellani}}, \bibinfo {author}
  {\bibfnamefont {A.~M.}\ \bibnamefont {García-García}},\ and\ \bibinfo
  {author} {\bibfnamefont {L.}~\bibnamefont {Benfatto}},\ }\href
  {https://doi.org/10.1103/PhysRevB.96.094514} {\bibfield  {journal} {\bibinfo
  {journal} {Phys. Rev. B}\ }\textbf {\bibinfo {volume} {96}},\ \bibinfo
  {pages} {094514} (\bibinfo {year} {2017})}\BibitemShut {NoStop}%
\bibitem [{\citenamefont {Deutscher}\ \emph
  {et~al.}(1973{\natexlab{b}})\citenamefont {Deutscher}, \citenamefont
  {Gershenson}, \citenamefont {Grünbaum},\ and\ \citenamefont
  {Imry}}]{Deutscher1973a}%
  \BibitemOpen
  \bibfield  {author} {\bibinfo {author} {\bibfnamefont {G.}~\bibnamefont
  {Deutscher}}, \bibinfo {author} {\bibfnamefont {M.}~\bibnamefont
  {Gershenson}}, \bibinfo {author} {\bibfnamefont {E.}~\bibnamefont
  {Grünbaum}},\ and\ \bibinfo {author} {\bibfnamefont {Y.}~\bibnamefont
  {Imry}},\ }\href {https://doi.org/10.1116/1.1318416} {\bibfield  {journal}
  {\bibinfo  {journal} {Journal of Vacuum Science and Technology}\ }\textbf
  {\bibinfo {volume} {10}},\ \bibinfo {pages} {697} (\bibinfo {year}
  {1973}{\natexlab{b}})}\BibitemShut {NoStop}%
\bibitem [{\citenamefont {Koyama}(2004)}]{Koyama2004}%
  \BibitemOpen
  \bibfield  {author} {\bibinfo {author} {\bibfnamefont {T.}~\bibnamefont
  {Koyama}},\ }\href {https://doi.org/10.1103/PhysRevB.70.226503} {\bibfield
  {journal} {\bibinfo  {journal} {Phys. Rev. B}\ }\textbf {\bibinfo {volume}
  {70}},\ \bibinfo {pages} {226503} (\bibinfo {year} {2004})}\BibitemShut
  {NoStop}%
\bibitem [{\citenamefont {Deutscher}(2021)}]{Deutscher2021}%
  \BibitemOpen
  \bibfield  {author} {\bibinfo {author} {\bibfnamefont {G.}~\bibnamefont
  {Deutscher}},\ }\href {https://doi.org/10.1007/S10948-020-05773-Y/FIGURES/1}
  {\bibfield  {journal} {\bibinfo  {journal} {Journal of Superconductivity and
  Novel Magnetism}\ }\textbf {\bibinfo {volume} {34}},\ \bibinfo {pages} {1699}
  (\bibinfo {year} {2021})}\BibitemShut {NoStop}%
\bibitem [{\citenamefont {Bandyopadhyay}\ \emph {et~al.}(1982)\citenamefont
  {Bandyopadhyay}, \citenamefont {Lindenfeld}, \citenamefont {McLean},\ and\
  \citenamefont {Sin}}]{Bandyopadhyay1982}%
  \BibitemOpen
  \bibfield  {author} {\bibinfo {author} {\bibfnamefont {B.}~\bibnamefont
  {Bandyopadhyay}}, \bibinfo {author} {\bibfnamefont {P.}~\bibnamefont
  {Lindenfeld}}, \bibinfo {author} {\bibfnamefont {W.~L.}\ \bibnamefont
  {McLean}},\ and\ \bibinfo {author} {\bibfnamefont {H.~K.}\ \bibnamefont
  {Sin}},\ }\href {https://doi.org/10.1103/PhysRevB.26.3476} {\bibfield
  {journal} {\bibinfo  {journal} {Phys. Rev. B}\ }\textbf {\bibinfo {volume}
  {26}},\ \bibinfo {pages} {3476} (\bibinfo {year} {1982})}\BibitemShut
  {NoStop}%
\bibitem [{\citenamefont {Devoret}\ \emph {et~al.}(2007)\citenamefont
  {Devoret}, \citenamefont {Girvin},\ and\ \citenamefont
  {Schoelkopf}}]{Devoret2007}%
  \BibitemOpen
  \bibfield  {author} {\bibinfo {author} {\bibfnamefont {M.}~\bibnamefont
  {Devoret}}, \bibinfo {author} {\bibfnamefont {S.}~\bibnamefont {Girvin}},\
  and\ \bibinfo {author} {\bibfnamefont {R.}~\bibnamefont {Schoelkopf}},\
  }\href {https://doi.org/10.1002/andp.200751910-1109} {\bibfield  {journal}
  {\bibinfo  {journal} {Annalen der Physik}\ }\textbf {\bibinfo {volume}
  {519}},\ \bibinfo {pages} {767} (\bibinfo {year} {2007})}\BibitemShut
  {NoStop}%
\end{thebibliography}%

\newpage
\onecolumngrid

\setcounter{figure}{0}
\renewcommand{\figurename}{FIG.}
\renewcommand{\thefigure}{S\arabic{figure}}

\setcounter{table}{0}
\renewcommand{\tablename}{TAB.}
\renewcommand{\thetable}{S\arabic{table}}

\setcounter{equation}{0}
\renewcommand{\theequation}{S\arabic{equation}}

\section{Supplementary material}

\subsection{Measurement setup and sample holder design}

\begin{figure*}[h]
	\includegraphics[width=\textwidth]{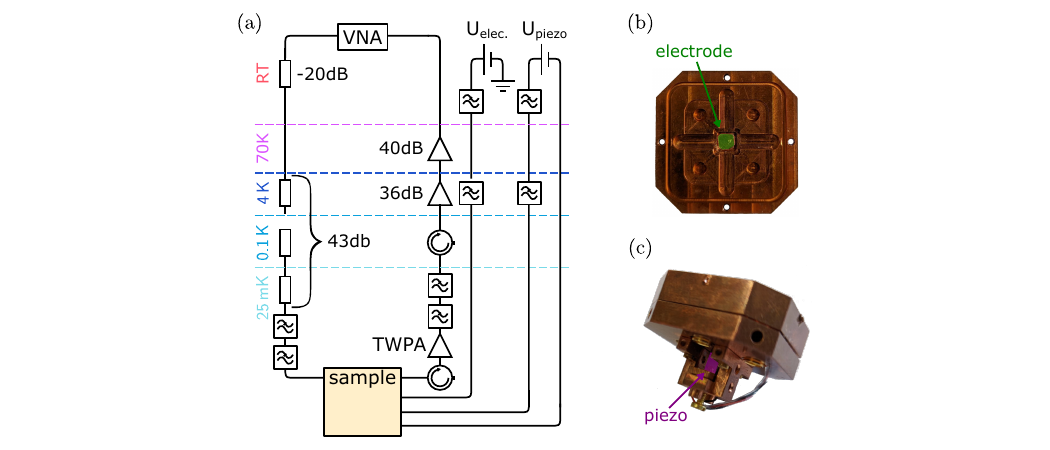}
	\caption{Measurement setup and sample holder. \textbf{(a)} Schematic of the experimental setup and wiring. \textbf{(b)+\textbf{(c)}} Photograph of the sample holder highlighting the DC electrode \textbf{(b)} and the piezo-actuator \textbf{(c)}.}   
	\label{Fig1_sup}
\end{figure*}

A schematic of the measurement setup is depicted in Fig.~\ref{Fig1_sup}(a). A commercial vector network analyzer (VNA) is used to probe the complex transmission $S_{21}$ of the sample. The input signal is attenuated by $-10 \, \textrm{dB}$ at multiple temperature stages of a dry dilution refrigerator and then filtered through home-made infrared plus commercial low-pass filters before reaching the sample holder. There, a SMP connector feeds the signal to a microstrip PCB which is wire-bonded to the on-chip transmission line. The output signal is amplified at base temperature by a superconducting travelling wave parametric amplifier and two high-electron-mobility transistors (HEMT) mounted at $4 \, \textrm{K}$ and $70 \, \textrm{K}$. 

\vspace*{0.2cm}
\vspace*{0.2cm}

As described in the main text, a piezo-actuator and a DC electrode are used to tune the TLS frequency \cite{Grabovskij2012, Lisenfeld2019}. The former is fixated at the bottom of the sample holder by a custom made, H-shaped mount with a center screw that allows making tight contact between the piezo and the sample backside (Fig.~\ref{Fig1_sup}(b)) The DC electrode is placed in a pocket in the sample holder lid, glued to the center pin of a SMA connector (Fig.~\ref{Fig1_sup}(c)). The necessary control voltages $U_{\rm piezo}$ and $U_{\rm elec}$ are generated by two independent sources at room temperature. For the DC lines, appropriate low-pass filters are installed at different temperature stages.

\newpage
\subsection{Resonator-TLS interaction and reference measurement}
While strongly coupled TLS ($g \approx \kappa \approx 1 \, \textrm{MHz}$) are the focus of the analysis presented in the main text, they only make up for $\sim 1 \%$ of all observed TLS (see Table \ref{tab1}). The majority of TLS are only moderately coupled ($g \ll 1 \, \textrm{MHz}$) to the resonator, yielding less pronounced anti-crossings. Sometimes, the symmetry point of a TLS is close to $\omega_\textrm{r}$. If this TLS also has a low loss rate ($\gamma_{TLS} \lesssim 1 \, \textrm{MHz}$), the trace of the hyperbolic TLS spectrum $\omega_{\rm TLS} = \frac{1}{\hbar}\sqrt{\Delta^2+\left(\epsilon + 2dE_\mathrm{z} \right)^2}$ is directly visible in the transmission data. 

\begin{figure*}[h]
	\includegraphics[width=\textwidth]{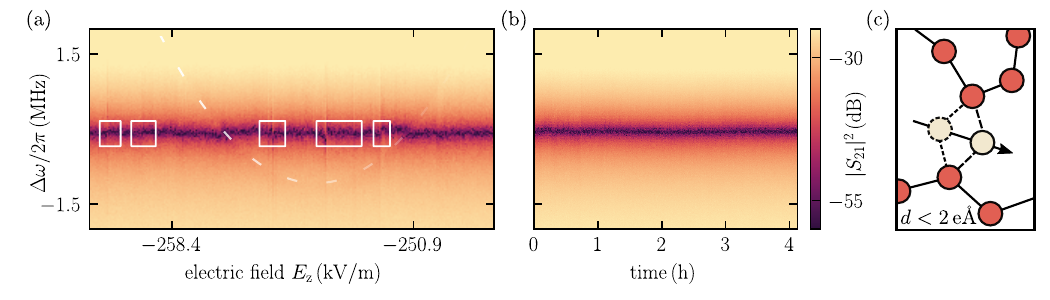}
	\caption{Electric field tuning of TLS. \textbf{(a)} Resonator transmission as a function of the applied electric field $E_\mathrm{z}$. White boxes indicate anti-crossings with moderately coupled TLS. The fading, dashed line highlights the trace of (visible) TLS with $d=1.0 \textrm{e\AA }$. \textbf{(b)} Reference measurements without applied fields or strain of the same resonator \textbf{(c)} Illustration of a atomic defect tunneling between two nearly degenerate configurations. The corresponding dipole is expected to not surpass $2 \textrm{e\AA }$.}    
	\label{Fig2_sup}
\end{figure*}

Such a TLS-resonator interaction is shown in Figure \ref{Fig2_sup}(a), where the white boxes highlight anti-crossings and the fading, dashed line is a fit to $\omega_{\rm TLS}$. The dipole moments extracted from such fits are typically on the order of $d \approx 1 \textrm{e\AA }$, in good agreement with a model of atomic defects (Fig.~\ref{Fig2_sup}(c)) in the natural oxide layer at the resonator surface \cite{Muller2019}. 

Figure~\ref{Fig2_sup}(b) shows a reference measurement of the same resonator (E1), recorded for several hours. Without an applied electrical field or strain, we do not observe any systematic resonator shift.

\vspace*{0.2cm}
\vspace*{0.2cm}

\newpage
\subsection{Resonator and film parameters}
\begin{figure*}[h]
	\includegraphics[width=\textwidth]{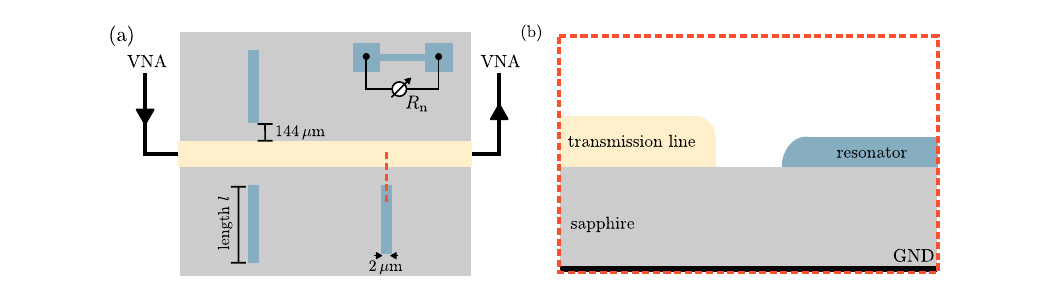}
	\caption{Sketch of resonator circuit design (not to scale). \textbf{(a)} Chip layout after structuring the films. \textbf{(b)} Cross section of the chip.}  
	\label{Fig3_sup}
\end{figure*}

All samples are fabricated from a total of five granular aluminum films, A to E, each sputter-deposited on a $430\,\si{\micro\meter}$ thick sapphire waver using an aluminum target and an argon plasma. These films predominantly differ in their normal state sheet resistance $R_\mathrm{n}$, which is tuned by the oxygen partial pressure during the sputtering process (see Ref.~\cite{Rotzinger2017} for details). Due to the sensitivity of $R_\mathrm{n}$ on the deposition conditions, it is monitored in-situ during the deposition, together with the thickness $t$ of the films \cite{Wildermuth2022}. Granular films are inherently inhomogeneous and we expect that the local variations average out over a distance of a few grain-sizes. On a larger scale, we observe a variation on the order of about 10\,\% over one wafer size (20\,mm $\times$ 20\,mm) measuring the sheet resistance of on-chip test structures (dumbbell structure Fig.~\ref{Fig3_sup}a)). 

Using conventional optical lithography techniques, the films are structured into microstrip microwave circuit with a layout as sketched in Fig.~\ref{Fig3_sup}a). Each chip contains multiple $\lambda/2$ resonators with varying length $l$ (see Table~\ref{tab1}) at a width of $2\,\si{\micro\meter}$, separated from a transmission line by $144\,\si{\micro\meter}$ ($0.5\,\mathrm{fF}$ coupling capacitance). We also estimate the sheet resistance value of each resonator by comparing the measured resonator frequencies $\omega_{\rm r}$ with simulations done in a 3D planar high-frequency electromagnetic software (SONNET). The respective values are listed in Table~\ref{tab1}. The effect of an dielectric capping is investigated by covering the resonators on chip D and E with an additional, insulating ($R_\mathrm{n} \gg R_\mathrm{q}$) granular aluminum film during the initial deposition step. The sheet resistance of these layers is estimated from the in-situ resistance measurement performed during the deposition.

\vspace*{0.2cm}
\vspace*{0.2cm}

From the count of all visible anti-crossings and hyperbolas, normalized to the total range $\Delta E_\mathrm{z}=1258 \, \mathrm{kV/m}$ of the electric field sweep, one obtains a rough estimate for the total density of observable TLS $D_{\rm E}$ for each resonator. These densities are two orders of magnitudes larger than the respective densities $D_{\rm E,g}$ including only the strongly coupled TLS. However, $D_{\rm E}$ and $D_{\rm E,g}$ depend on the experimental implementation of the DC electrode and do not take into account that some TLS hyperbolas are more likely to be observed. A better estimate is provided by the spectral TLS density, which can be calculated as follows

\begin{align}
	D_{\omega} = \sum_\textbf{i} \frac{\Delta E_\mathrm{z}}{d_\textrm{i}^*}.
	\label{EqS1}
\end{align}

Here, i is the index over all observed TLS and $d_\textrm{i}^*$ is their individual slope with respect to the electric field, estimated from the measurement data. The corresponding values are given in Tab.~\ref{tab1}. The average spectral density of the strongly coupled TLS is $\overline{D}_{\omega,\rm g} \approx 0.01 \, \rm GHz^{-1}$. Note that counting of moderately coupled TLS was not possible on resonators C1-C3, where most anti-crossings are small compared to the pronounced resonance frequency fluctuations measured on this chip \cite{Kristen2023}.  

The experimentally relevant properties of the measured two dimensional ($\xi \approx 10 \, \textrm{nm} \sim t$ \cite{Voss2021}) superconducting granular aluminum films generally depend on their resistivity $\rho = R_\mathrm{n}t$. For example. an estimate for the London penetration depth at low temperatures $T \ll T_\textrm{c}(\rho) \approx 1.8 \, \textrm{K}$ is given by $\lambda_\textrm{L} \approx 1.05 \times 10^{-3} \sqrt{\rho/2} = 4.5 \pm 2.1 \, \si{\micro\metre}$ \cite{Rotzinger2017}. Consequently, the films ($\textrm{t} \ll \lambda_\textrm{L}$) are transparent for AC fields. 

An estimate of the penetration depth of the DC electric fields is challenged by the heterogeneous granular structure of the film, which consists of both metallic and dielectric parts. However, a lower limit can be stated by calculating the Thomas-Fermi length of a clean superconductor with similar properties. This yields $\lambda_\textrm{TF} \approx \sqrt{2\lambda_\textrm{L}^2 E_\textrm{F}^*/(3c^2m_\textrm{m}^*)} = 2.1 \pm 0.1 \, \si{\nano\metre}$ \cite{Koyama2004}, which is on the order of the grain size. Here $E_\textrm{F}^* \propto \rho^{-0.7}$ and $m_\textrm{e}^* \propto \rho^{-0.44}$ are the effective Fermi energy and charge carrier mass, respectively \cite{Bachar2015}. From $\lambda_\textrm{L}$, the superconducting charge carrier density of the granular films $n_0 = m_\textrm{e}^*/(2\mu_0e^2\lambda_\textrm{L}^2)$ can be estimated, which is about three orders of magnitudes smaller then for pure aluminum. 

One of the few available indicators for the degree of disorder in a thin film is the Ioffe-Regel parameter $k_\textrm{F}l = \hbar (3 \pi^2)^{2/3}/(e^2\rho n^{1/3})$. Following Ref.~\cite{Deutscher2021}, superconductivity in granular aluminum is destroyed due to disorder for a critical value of $(k_\textrm{F}l)_\textrm{c} \sim 0.01$. For materials where disorder predominantly occurs on the atomic scale $(k_\textrm{F}l)_\textrm{c} \approx 1$. Using $n=1/(\mu_\textrm{H} \rho e)$, where $\mu_\textrm{H} \propto \rho^{-0.5}$ is the Hall mobility \cite{Bandyopadhyay1982}, yields values ranging from $0.14$ to $0.8$, i.e., all films are highly disordered and one can expect mechanisms related to the superconductor to insulator transition to become relevant.

\vspace*{0.2cm}
\begin{table}[h]
	\caption{\label{tab1}Characteristics of the measured resonators, including their resonance frequencies $\omega_{\rm r}$, linewidths $\kappa$ ($\overline{n} = 100$), sheet resistances $R_\mathrm{n}$ and kinetic sheet inductances $L_\mathrm{kin}=0.18 \hbar R_\mathrm{n}/(T_\textrm{c} k_\textrm{B})$ \cite{Rotzinger2017} as well as physical dimensions (length: $l$, thickness: $t$, width: $2\,\si{\micro\meter}$). The TLS densities with respect to the electric field ($D_\mathrm{E}$) or frequency spectrum ($D_\mathrm{\omega}$) are estimated as described in the main text. The subscript $g$ indicates strongly coupled TLS.}
	\begin{ruledtabular}
		\begin{tabular}{ccccccccccc}
			Reso. & $\omega_{\rm r}/2\pi \, {\rm (GHz)}$ & $\kappa/2\pi \, {\rm (MHz)}$  & l\,($\si{\micro\meter}$) & t\,(nm) & $R_\mathrm{n} \, ({\rm k\Omega})$ & $L_\mathrm{kin} \, ({\rm nH})$  & $D_\textrm{E} \, \textrm{(TLS/kV)}$ & $D_\omega \, \textrm{(TLS/GHz)}$ & $D_{\rm E,g} \, \textrm{(gTLS/kV)}$ \footnotemark[1]  \\ \hline
			A1 & 10.565 & 2.33 & 406 & 25 & 0.59 & 0.45  &  $\sim 140 $ & $\sim 21$ & 1.6 \\
			A2 & 10.740 & 2.69 & 390 & 25 & 0.61 & 0.47  &  - & - & 0.4\\
			B1 & 5.494  & 1.84 & 505 & 22 & 1.39 & 1.05  &  $\sim 290$ & $\sim 34$ & 2.0 \\
			B2 & 6.154 & 2.29 &  440 & 22 & 1.49 & 1.14  &  $\sim 250$ & $\sim 27$ & 2.4 \\
			B3 & 6.793 & 2.61 & 390 & 22 & 1.53 & 1.17  & - & - & 2.0 \\
			C1 & 4.069 & 0.61 & 406 & 30 & 3.97 & 3.03  &  - & - & 4.0 \\
			C2 & 4.663 & 0.82 & 337 & 30 & 4.32 & 3.30  &  - & - & 4.8 \\
			C3 & 5.780 & 1.51 & 287 & 30 & 3.76 & 2.87  &  - & - & 4.0 \\
			D1 & 8.006 & 1.72 & 505 & 24+18 & 0.65+12 & 0.5    & $\sim 220$ & $\sim 24$ & 2.4 \\
			D2 & 9.010 & 2.95 & 440 & 24+18 & 0.69+12 & 0.53   & $\sim 260$ & $\sim 43$ & 1.6 \\
			D3 & 9.995 & 2.95 & 390 & 24+18 & 0.71+12 & 0.54  & - & - & 2.0 \\
			E1 & 7.839 & 1.56 & 505 & 23+17 & 0.68+190 & 0.52  & $\sim 450$ & $\sim 29$ &4.4 \\
			E2 & 8.872 & 0.44 & 440 & 23+17 & 0.72+190 & 0.55  & $\sim 410$ & $\sim 33$ & 3.2 \\	
		\end{tabular}
	\end{ruledtabular}
	\footnotetext[1]{Average value from two measurements performed over two independent cooldowns ($\Delta E_\mathrm{z, tot}=2 \times 1258 \, \mathrm{kV/m}$).}
\end{table}

\newpage
\subsection{Electric field simulations}

\vspace*{0.2cm}
\begin{figure*}[b]
	\includegraphics[width=\textwidth]{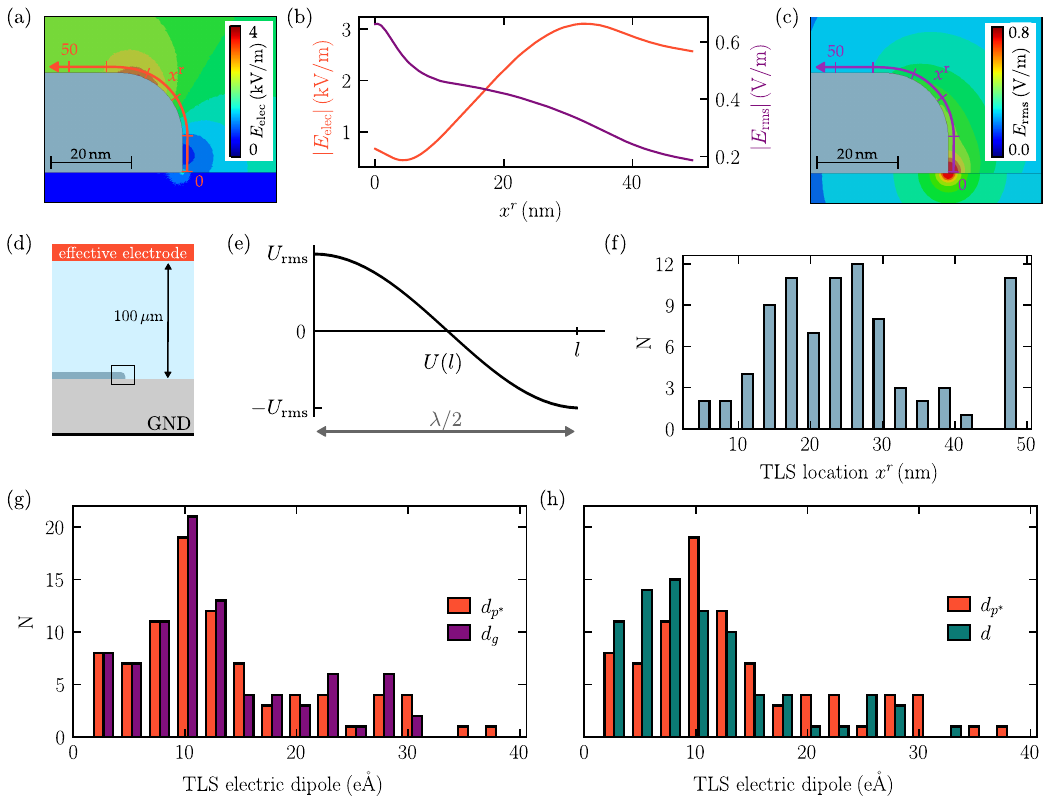}
	\caption{Resolving the dipole moment and position of the observed TLS. \textbf{(a)} Simulated electric field distribution around the resonator edge for $U_{\rm elec} = 1 \, \textrm{V}$ ($q_\mathrm{reso}=const.$) \textbf{(b)} Absolute values of the electric field from the DC electrode (${E}_{\rm elec}$) and the resonator (${E}_{\rm rms}$) along the film edge $x^\mathrm{r}$. \textbf{(c)} Simulated electric field distribution around the resonator edge for $U_\textrm{rms}= 1 \, \si{\micro\volt}$ ($U_{\rm elec} = 0 \, \textrm{V}$) \textbf{(d)} Sketch of the 2D model (not to scale) used for the simulation \textbf{(e)} Distribution of the effective voltage for the fundamental mode of a $\lambda/2$-resonator with lenght $l$. \textbf{(f)} Histogram of the TLS positions along the film edge obtained from solving Eq.~\eqref{EqS4}. \textbf{(g)} Comparison between the distributions of the TLS dipole moments calculated from the slope of the TLS hyperbola ($d_{\rm p^*}$) and the coupling strength to the resonator ($d_\mathrm{g}$). \textbf{(h)} Dipole distributions obtained from the slope of the TLS hyperbola $p^*$, using either the location dependent electric field value $\bm{E}_{\rm elec}(\bm{x^r})$ ($d_{\rm p^*}$) or the maximum field $E_{\rm z}$ ($d$).}
	\label{Fig4_sup}
\end{figure*}
In order to estimate the TLS dipole moment from the slope of the fitted anticrossings $p^*$, knowledge of the electric field generated by the DC electrode ($\bm{E}_{\rm elec}(U_{\rm elec},\bm{x})$) at $\bm{x}$, the position of the TLS, is necessary. Therefore, simulations with an electrostatic finite element solver (ANSYS Maxwell) were performed. Figure \ref{Fig4_sup}(d) shows a sketch of the 2D model employed in these simulations, including an effective DC electrode which takes into account the actual field distribution ($U_{\rm elec}^{eff} \approx 0.3U_{\rm elec}$) extracted from a (computationally expensive) 3D simulation. For all simulations, the granular aluminum film was modeled as a perfect conductor.

Figure \ref{Fig4_sup}(a) shows the simulated distribution of the electric field produced by the DC electrode for $U_{\rm elec} = 1 \, \textrm{V}$. The corresponding absolute field values at a positional axis $x^r$ along the film edge are plotted in Fig.~\ref{Fig4_sup}(b) (orange line). Because the exact position of the TLS is generally not known, the maximum field $E_{\rm z}=|\bm{E}_{\rm elec}(\bm{x^r})_\mathrm{max}|= 3144 \, m^{-1} \cdot U_{\rm elec}$ value along $x^r$ is used to approximate the conversion $p^*U_{\rm elec} \rightarrow 2dE_\mathrm{z}/\hbar$. Here $p^*$ is the slope of the TLS hyperbola with respect to the applied voltage measured in the experiment. Thus,
\begin{align}
	d = \frac{ U_{\rm elec} \cdot \hbar p^*}{2 E_{\rm z}} = \frac{\hbar p^*}{6288 \, m^{-1}}.
	\label{Eq_d}
\end{align} 
This formula is used to calculated the lower bounds to the TLS dipole moments discussed in the main text. 

Here, we show an alternative approach for determining the TLS dipole moment, which makes use of the measured coupling rates $g$ between the TLS and the resonator modes. There, we are only interested in the fundamental mode, corresponding to half a wavelenght $(\lambda/2)$ for our resonator geometry (see Fig.~\ref{Fig4_sup}(e)). The voltage will be maximal at the two ends of the resonator, where its effective value is equal to the single photon root mean square voltage $U_\textrm{rms}=\sqrt{h^2 \omega_{\rm r}^2 Z/(8\pi e^2 \alpha Z_\textrm{vac})} \approx 15 \, \si{\micro\volt}$ \cite{Devoret2007}. Here, $Z_\textrm{vac}$ is the vacuum impedance, $\alpha$ the fine structure constant and $Z=\omega_{\rm r} L_\mathrm{n} l/w$ is the impedance of the $w = 2 \, \si{\micro\meter}$ wide resonator stripe. For the simulations of the corresponding field of the resonator $\bm{E}_{\rm rms}(U_{\rm rms},\bm{x})$, we raise the potential of the resonator edge to $U_\textrm{rms}$, while $U_\mathrm{elec}$ is set to zero. The resulting distribution of $E_\mathrm{rms}$ is shown in Fig.~\ref{Fig4_sup}(c), and the absolute values along $x^r$ are shown in Fig.~\ref{Fig4_sup}(b) (purple line).

Considering the full distribution of electric fields along the film edge $x^r$, there are two separate equations which allow calculation of the TLS dipole component parallel to the respective electric field

\begin{gather}
	d_{\rm p^*} = \frac{\mathrm{1V}\cdot\hbar p^*}{2|\bm{E}_\textrm{elec}(U_{\rm elec}=1 \, \textrm{V}, x^r)|} \label{EqS2}, \\
	d_\mathrm{g} = \frac{\omega_{TLS}}{\Delta}\frac{\mathrm{1V}\cdot\hbar^2 g}{|\bm{E}_\textrm{rms}(U_{\rm rms}, x^r)|} \approx \frac{\mathrm{1V}\cdot\hbar g}{|\bm{E}_\textrm{rms}(U_{\rm rms}, x^r)|} \label{EqS3}.
\end{gather}

In Eq.~\eqref{EqS3} we assume $\Delta \gg \epsilon$ for the TLS energies, since all TLS are evaluated near the symmetry point. Because the position of each individual TLS $x^r$ is not known, these equations are generally not applicable. However, it is reasonable to assume that $\bm{E}_\textrm{rms} \parallel \bm{E}_\textrm{elec}$ at the film surface, which means $d_{\rm g}(x^r)$ and $d_{\rm p^*}(x^r)$ should generally be identical. Thus, one can explicitly solve this set of equations for $x^r$ by minimizing

\begin{align}
	\min_{x^r \in [0,50]} |d_\textrm{g}(x^r)-d_{\rm p^*}(x^r)|.
	\label{EqS4}
\end{align} 

Figure~\ref{Fig4_sup}(f) shows the resulting distribution of $x^r$ for all 86 fitted anti-crossings, revealing the possible TLS locations along the film edge. Inserting these positions into Eq.~\eqref{EqS2} and Eq.~\eqref{EqS3} results in two distributions for the TLS dipole moments, calculated from the coupling to the DC electrode ($d_{\rm p^*}$) and from the coupling to the resonator ($d_{\rm g}$). The large (97,9\%) overlap of these distribution, which is illustrated by Fig.~\ref{Fig4_sup}(g), is remarkable. In particular, it supports the assumption that all observed TLS are located near the film-vacuum interface (see Ref.~\cite{Bilmes2020} for a more detailed discussion on the prevalence of TLS at different interfaces) as well as the estimated shape of the resonator corner. While the latter has its origin in the slightly isotropic dry-etching process, we note that the obtained results do not change significantly when a sharp resonator edge is assumed instead. Even directly at such a sharp corner, the field only reaches a value of 2$E_\mathrm{z}$ ($d \rightarrow d/2$), resulting in an average TLS dipole moment that is still extremely large. However, nanoscopic variations in the film structure of this kind might explain the discrepancies between $d_{\rm g}$ and $d_{\rm p^*}$ observed at large dipole values.

Figure.~\ref{Fig4_sup}(h) compares $d_{\rm p^*}$ to the distirbution obtained from Eq.~\ref{Eq_d}, which have an reasonable overlap of 74,7\%. Like Fig.~\ref{Fig4_sup}(f), it is obtained under the assumption that TLS in resonators covered with an additional layer (chip D and E) are subjected to a reduced resonator field, because spatially they are further separated from the charges contributing to $\bm{E}_\textrm{elec}$. This assumption is in accordance with the data for $g$ presented in the main text (Fig.~3(b)). Since properly simulating the granular, insulating top layer is challenging, a reduction of the resonator fields near the surface of the two-layer film according to $|E_\mathrm{rms}|/n$ is empirically assumed instead. We find that the overlap between the different dipole distributions is largest for $n=2$.

\newpage
\subsection{Fitting model and routine}

For the quantitative analysis of the measurement data, it is sufficient to consider a resonator coupled to a single TLS with strength $g$ and to a transmission line with strength $\Omega$. The transmission line is terminated by impedance mismatched bonding wires on both ends, and therefore best modeled as an additional cavity \cite{Khalil2012}. The full Hamiltonian of such a system reads

\begin{align}
	\hat{\mathcal{H}}_\textrm{sys}/\hbar = \omega_{\rm t} \hat{t}^\dagger\hat{t}+ \omega_{\rm r} \hat{r}^\dagger\hat{r} + \Omega(\hat{t}^\dagger\hat{r}+\hat{r}^\dagger\hat{t})-
	ig(\hat{\sigma}_i\hat{r}^\dagger-\hat{\sigma}_+\hat{r})+\omega_{\rm TLS}\sigma_\textrm{z},
	\label{EqS5}
\end{align}

where $\hat{t} (\hat{r})$ is the photon annihilation operators of the transmission line (resonator) and
$\hbar\omega_{\rm TLS} = \sqrt{\Delta^2+\left(\epsilon + 2 d |E_{\rm elec}|\right)^2}$ is the TLS transition energy. Applying input-output theory to Eq.~\eqref{EqS5} yields \cite{Sarabi2015}

\begin{align}
	S_{21}(\omega,E_\textrm{elec}) = ae^{-i\alpha}\left( 1-\frac{\overline{\kappa}_{\rm c}/2}{\frac{\kappa}{2}+i(\omega-\omega_{\rm r})+\frac{g^2}{\frac{\gamma_{TLS}}{2}+i(\omega-\omega_{TLS})}} \right)
	\label{EqS6}
\end{align}

for the transmission in the low temperature and low power limit ($k_\textrm{B}T \ll \hbar \omega$ and $ \overline{n} \ll 1$). In Eq.~\eqref{EqS6}, $\kappa$ and $\gamma_{TLS}$ describe the loss of the resonator and the TLS, respectively, e.g. to the bath of weakly coupled TLS. The small complex component of the resonator's external loss rate $\overline{\kappa}_{\rm c}=\kappa_{\rm c}e^{i\phi(\Omega)}$ includes the asymmetry of the resonator signal induced by the transmission line mode. To account for the net attenuation/amplification of the experimental setup and uncorrected phase delay, it is useful to introduce two additional parameters $a$ and $\alpha$.

\begin{figure*}[h]
	\includegraphics[width=\textwidth]{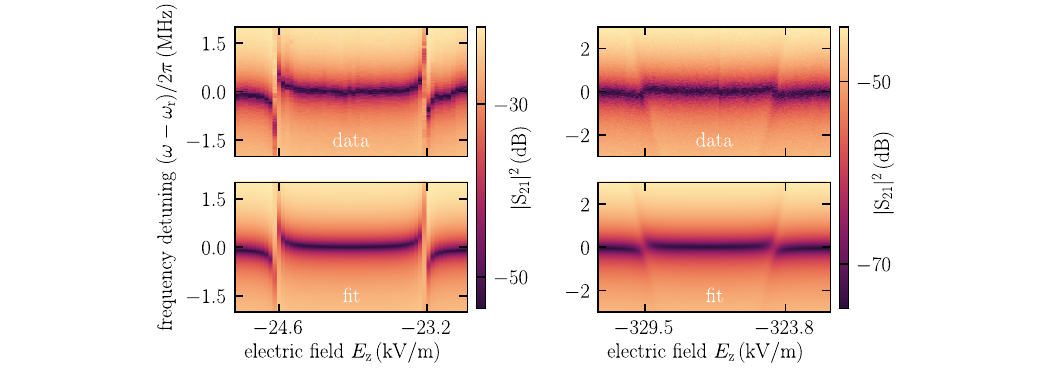}
	\caption{Anti-crossing fits. Top panel: Transmission amplitude of resonator E1 (left) and A1 (right) showing the interaction with a single strongly coupled TLS that is tuned through resonance by the applied electric field $E_\mathrm{z}$. Note that on the right, the trace of the TLS is directly visible in the spectroscopy data. Bottom panel: Fit of Eq.~\eqref{EqS6} to the experimental data.}    
	\label{Fig5_sup}
\end{figure*}

The fitting routine is as follows: First, a reduced model $|S_{21}(\omega, g=0)|$ is fitted to the bare transmission data to extract the parameters of a given resonator ($a,\alpha,\kappa_{\rm c},\phi,\kappa,\omega_{\rm r}$). These parameters as treated as constants in the second step, where the full model $|S_{21}(\omega,E_\textrm{elec})|$ is separately fitted to each pronounced anti-crossing observed in that resonator, extracting the corresponding TLS parameters ($g,\gamma_{TLS},\Delta, \epsilon,d$). Two examples illustrating the results of this fitting routine are shown in Fig.~\ref{Fig5_sup}.

\vspace*{0.2cm}
\vspace*{0.2cm}

\newpage
\subsection{Loss tangent}

\begin{figure*}[h]
	\includegraphics[width=\textwidth]{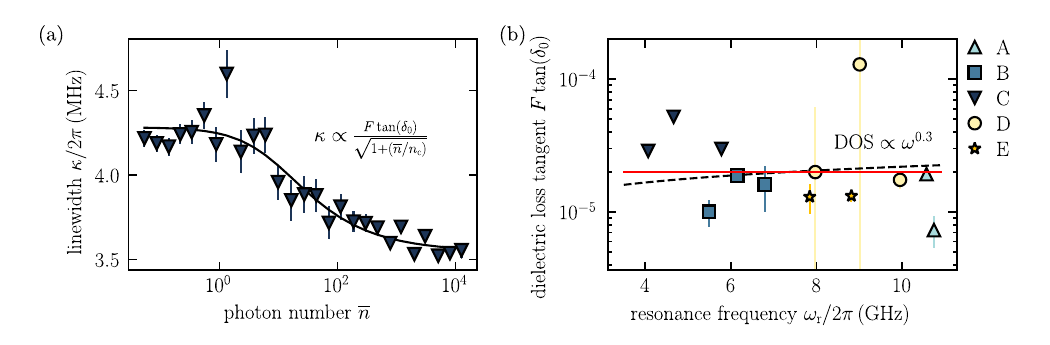}
	\caption{Resonator loss. \textbf{(a)} Linewidth $\kappa$ of resonator C1 as a function of the average circulating photon number. Solid line is a Fit to Eq. \eqref{EqS7}. \textbf{(b)} Fitted loss tangents toghether with the expected scaling $\propto\omega^\mu$ due to interacting TLS (dashed black line, $\mu = 0.3$) and the standard tunneling model result (solid red line, $\mu = 0$). Error bars smaller than the marker size are not shown.}    
	\label{Fig6_sup}
\end{figure*}

To investigate whether the main loss channel of our resonators originates from a large ensemble of weakly coupled TLS, we measure the loss rate $\kappa$ as a function of the the average circulating photon number $\overline{n}$. In the scope of the standard tunneling model for TLS \cite{Anderson1972, Phillips1972}, the loss rate is typically expressed as

\begin{align}
	\frac{\kappa}{\omega_{\rm r}} = \frac{F \tan(\delta_0)}{\sqrt{1+\overline{n}/n_\textrm{c}}} + C
	\label{EqS7}
\end{align}

where $n_\textrm{c}$ is the critical photon number above which resonant TLS start to saturate and $F \tan(\delta_0)$ is the dielectric loss tangent of our film dressed by a filling factor F. The measured data for resonator C3 and a fit to Eq. \eqref{EqS7} is shown in Fig.~\ref{Fig6_sup}(a). We find $F \tan(\delta_0) = 3.2 \pm 0.1 \times 10^{-5}$. Assuming that the majority of TLS reside in a $4 \, \textrm{nm}$ thick, amorphous oxide layer on the resonator surface allows an estimate for the density of weakly coupled TLS $D_\Delta = 3 h \epsilon_0 \epsilon_r \tan(\delta_0)/(\pi d^2) \approx 1.3 \times 10^6 \, \rm GHz^{-1}$, much larger than the densities found for moderately and strongly coupled TLS (Tab. \ref{tab1}). Here, $F \approx 10^{-4}$, $d = 0.5 \, \rm e\AA$ and $\epsilon_r \approx 10$ for $\rm AlO_x$ were used. 

The fitted loss tangents of all resonators are plotted in Fig.~\ref{Fig6_sup}(b). The corresponding quality factors $Q_\mathrm{i} = 1/F \tan(\delta_0)$ on the order of $10^5$ are typical for granular aluminum films \cite{Rotzinger2017, Grunhaupt2018}. Due to the scatter of the data, a potential scaling of the TLS density of state (DOS) can not be resolved \cite{Burnett2014, Kristen2023}.

\vspace*{0.2cm}
\vspace*{0.2cm}

\end{document}